\documentclass[%
amsmath,amssymb,
preprint,%
floatfix,
]{revtex4-1}

\usepackage{graphicx}% Include figure files
\usepackage{placeins}
\usepackage{dcolumn}% Align table columns on decimal point
\usepackage{bm}% bold math
\usepackage[utf8]{inputenc}
\usepackage[T1]{fontenc}
\usepackage{mathptmx}
\usepackage{xcolor}
\begin{document}
%	\title[]{Formation and evolution of various coherent structures in strongly coupled dusty plasmas using visco-elastic fluid description}
%	\title[]{Kelvin-Helmholtz instability and tracer transport in rotating strongly coupled dusty plasma}
%		\title[]{Kelvin-Helmholtz instability in strongly coupled dusty plasma under rotational sheared flows and tracer  transport}
	\title[]{Kelvin-Helmholtz instability in strongly coupled dusty plasma with rotational shear flows and tracer transport}
 
%		 advection of passive tracers with the flow}
		%: The interplay between the shear waves and the vortices of instability}	

	\author{Vikram Dharodi}%
	\email{dharodiv@msu.edu}
%	\email{vikram.dharodi@ipr.res.in}
		\affiliation{Institute for Plasma Research, HBNI, Bhat, Gandhinagar 382 428, India}%
	\author{Bhavesh Patel}
	\affiliation{Institute for Plasma Research, HBNI, Bhat, Gandhinagar 382 428, India}
	\author{Amita Das}%
	\affiliation{Department of Physics, Indian Institute of Technology, New Delhi, 110016, Delhi, India}%
	
	\date{\today}% It is always \today, today,
	\begin{abstract}
%%%%%%%%%%%%
Kelvin-Helmholtz (KH) instability plays a significant role in transport and mixing properties of any  medium. In this paper, we numerically explore this instability for a two-dimensional strongly coupled dusty plasma with rotational shear flows. We study this medium using generalized hydrodynamic fluid model which treats it as viscoelastic fluid. We consider the specific cases of rotating vorticity with abrupt radial profiles of rotation. In particular: single-circulation, and  multi-circulation vorticity shell profiles have been chosen. We observe  the KH vortices at each circular interface  between two relative rotating flows  along with a pair of ingoing and outgoing wavefronts of transverse shear waves. Our studies show that due to the interplay between KH vortices and shear waves  in  the strongly coupled medium, the mixing and transport behaviour are much better than inviscid hydrodynamic fluids. In interests of substantiating the mixing and transport behaviour, the generalized hydrodynamic fluid model  is extended to include the Lagrangian tracer particles. The numerical dispersion of these tracer particles in a flow provides an estimate of the diffusion in such a medium. We present the preliminary observations of tracers distribution (cluster formation) and their diffusion (mean square displacement) across the medium.
%
%To substantiate this observation,
%due to the existence of such waves in the strongly coupled medium, the mixing and transport behaviour in these fluids are much better than inviscid hydrodynamic fluids. 
%
%	the interplay between these shear waves and KH vortices results in better mixing of VE fluids than standard hydrodynamic fluids where the only interaction between KH vortices happens. 

	\end{abstract}
	%%%%%%%%%%%%%%%%%%%%%%%%%%%%%%%%%%%%%%%
	% for revtex4 here maketitle should be written 	
	\maketitle
	
%%%%%%%%%%%%%%%%%%%%%%%%%%%%%%%%%%%%%%%
%          INTRODUCTION
%%%%%%%%%%%%%%%%%%%%%%%%%%%%%%%%%%%%%%%

\section{Introduction} 
\label{introduction}
%~~~~~~~~~~~~~~~~~~~~~~~~~~~~~~
The Kelvin-Helmholtz (KH) instability has been ubiquitously observed in hydrodynamic fluids \cite{drazin1970kelvin,chandrhd1981hydro}, plasmas \cite{horton1987kelvin}, geophysical flows \cite{van2010deep}, and astrophysical situations \cite{foullon2011magnetic}. This instability occurs  in form of vortices at the interface between two flows in the presence of a velocity shear. The interactions between these KH vortices govern the transport processes like mixing and diffusion \cite{smyth2012ocean}. The main objectives of this study are to explore the formation and evolution of KH vortices of a rotating dusty plasma and quantifying the mixing they generate using the tracer particles simulation. The formation and evolution of these vortices depend on the shear and nature of the medium (here the  viscoelasticity of the system).   The KH instability for  dust flows has also been investigated theoretically  \cite{d1990kelvin,rawat1993kelvin,pandey2012shear,banerjee2012shear,dolai2016effect} and numerically \cite{ashwin2010kelvin,tiwari2012kelvina,tiwari2012kelvinb,tiwari2014kelvin} as well experimentally \cite{luo2001kelvin} for planar sheared flows.  The rotating vortex flows have been studied considerably \cite{konopka2000rigid,klindworth2000laser,schablinski2014sheared,choudhary2020three}, but to our knowledge, no prior studies have explicitly examined the KH instability for such flows, except ~\cite{dharodi2014visco, dharodi2020rotating}. Dharodi $et~al.$ ~\cite{dharodi2014visco} numerically studied  shearless smooth rotating flows to avoid KH destabilization and also considered sharp rotating  flows where KH arises in homogeneous medium. Dharodi  in \cite{dharodi2020rotating}  has explored these rotating sheared flows in heterogeneous medium.

A dusty plasma can exist in strong coupling state quite easily because of high charged dust particles, which is called strongly coupled dusty plasma (SCDP). The SCDP has been modelled under the formalism of generalized hydrodynamic (GHD) fluid model. This model treats the SCDP as a VE fluid and characterizes its VE effects  through the coupling strength parameter which is often measured as the ratio ${\eta/\tau_m}$, the coupling  parameters $\eta$ and $\tau_m$ are the shear viscosity and the Maxwell relaxation  time, respectively. {\it{Thus, here, the effects of viscoelasticity on KH vortices are observed by varying the ratio ${\eta/\tau_m}$}}. We consider the incompressible limit of GHD model which in addition to the evolution of hydrodynamic KH instability, also supports transverse shear (TS) waves that propagate at phase velocity $\sqrt{\eta/\tau_m}$ \cite{kaw1998low,dharodi2014visco,dharodi2016sub}. These propagating shear waves have the same symmetry as that of their source structure, until there is no boundary effect or no interaction with other waves or obstacle like vortex. Since our interest is in KH instability of rotating SCDPs, we consider the specific cases of sharp vorticity patches: (i) single-circulation, and (ii) multi-circulation vorticity shell profiles. The single-circulation case has already been somewhat discussed in~\cite{dharodi2014visco},  in the present paper we explore it in more detail.   We observe  the KH instability at each circular interface between two relative rotating flows in form of small vortices along with a pair of ingoing and outgoing wavefronts of TS waves. The interactions between interacting KH vortices and TS waves help the VE fluid in better mixing than standard HD fluids where the only interactions between the KH vortices take place. To substantiate this observation, the passive tracers are dispersed throughout the medium.

In the context of fluid mechanics the tracer transport has been studied extensively for flow visualization \cite{douady1991direct,fessler1994preferential} with the help of theoretical \cite{balkovsky2001intermittent,falkovich2004intermittent,zaichik2003two,bec2006acceleration}, computational \cite{squires1991preferential,boivin1998direct,reade2000numerical,zhou2001modelling,yeung2001lagrangian,yeung2002lagrangian,ishihara2002relative,collins2004reynolds,biferale2004multifractal,chun2005clustering,biferale2005particle,cencini2006dynamics} and experimental \cite{ott2000experimental,la2001fluid,mordant2001measurement,voth2002measurement,sawford2003conditional} approaches.  This technique is also used in complex fluids (polymers, colloids and biological materials) \cite{mason1997particle,waigh2005microrheology}. An analysis of the separation of the particle trajectories with the 2D hydrodynamic fluid is also being carried out by Falkovich et al. \cite{falkovich2001particles}. In dusty plasma, Schwabe et al. \cite{schwabe2014collective} observed the vortex movements by adding some micro particles around the void. Here, we consider two kinds of point-like tracer particles, (i) non-inertial, and (ii) inertial tracer particles. The tracer dynamics is simulated using a one-way coupled Lagrangian point-particle approach~\cite{riley1974diffusion,mclaughlin1989aerosol} which means the tracers are affected by the fluid flow, but not vice-versa. In case of multi-circulation vortex profiles, at intermediate time range, a complete picture of a turbulent flow is observed which has a collection of small KH vortices and waves. When the system is left for a very long time, it ultimately settles down to a single vortex faster than in HD fluid. It is observed that the relaxing rate of such turbulent medium increases with the increasing coupling strength. 

This paper has been organized as follows.  In section \ref{GHT2}, the GHD model especially developed for the study of SCDPs is extended to include the transport of passive Lagrangian tracers. This extended model is referred as incompressible Generalized Hydrodynamic Tracer Transport (i-GHTT) model.  Section \ref{simulation_metho} presents the numerical procedure in order to solve the set of equations of GHTT model. In Sec. \ref{results}, we numerically explore the evolution of different types of sharp rotating vorticities in VE fluids and quantifying the mixing they generate using the tracer particles simulations. Finally, Section~\ref{discussion} contains the discussion and the conclusions.

%%%%%%%%%%%%%%%%%%%%%%%%%%%%%%%%%%%%%%%
% Governing Equations
%%%%%%%%%%%%%%%%%%%%%%%%%%%%%%%%%%%%%%% 
\section{Generalized Hydrodynamic Tracer Transport (GHTT)  Model}
\label{GHT2}
%~~~~~~~~~~~~~~~~~~~~~~~~~~~~~~~
A dusty plasma can be prepared or found as a strongly coupled dusty plasma (SCDP) rather easily because of high charge on the micron-sized dust particles. Below the crystallization limit, a SCDP behaves like a viscoelastic fluid which favors both the incompressible transverse shear modes and the compressible longitudinal modes. To study such SCDP the generalized hydrodynamic (GHD) fluid model is found to be quite suitable which takes into account both types of modes. To scrutinize the effect of transverse modes and to abate the possible pairing with the longitudinal mode, we consider the incompressible limit of GHD (i-GHD) model. Thus, i-GHD model represents the incompressible SCDPs which support transverse modes only. The momentum and continuity equations for i-GHD of homogeneous strongly coupled  dusty plasma can be written as: 
%~~~~~~~~~~~~~~~~~~~~~~~~~~~~~~~~~~~~~~~
	\begin{equation}\label{eq:in_momentum}
	\left[1 + \tau_m \left(\frac{\partial}{\partial t}+\vec{v}_d  \cdot \nabla
		\right)\right]  \left[{\frac{\partial \vec{v}_d } {\partial t}+\vec{v}_d  \cdot
			\nabla\vec{v}_d } + \frac{\nabla P}{n_d} - {\nabla \phi_d} \right]  =  {\eta}{\nabla^2 }{\vec{v}_d}{,}
	\end{equation}
%~~~~~~~~~~~~~~~~~~~~~~~~~~~~~~~~~~~~~
and 
%~~~~~~~~~~~~~~~~~~~~~~~~~~~~~~~~~~~~~
	\begin{equation}\label{eq:in_continuity}
	\nabla \cdot \vec{v}_d= 0{,}
	\end{equation}
%~~~~~~~~~~~~~~~~~~~~~~~~~~~~~~~~~~~~~
respectively. $n_d$ is the number density which is normalized by its respective equilibrium value $n_{d0}$. The scalar  potential $\phi_d$ in the dusty plasma system is normalized by ${{K_B}{T_i}}/{e}$. The parameters $e$, $T_i$ and $K_B$ are the electronic charge,  ion temperature and Boltzmann constant, respectively.  The charge fluctuation over each dust grain has been ignored. The time and length are normalized by inverse of dust plasma frequency $\omega^{-1}_{pd}=\left({4\pi(Z_de)^{2}n_{d0}}/{m_{d0}}\right)^{-1/2}$, plasma Debye length $\lambda_{d}=\left({K_B T_i}/{4{\pi} {Z_d}{n_{d0}}{e^2}}\right)^{1/2}$, respectively. In incompressible limit the Poisson equation has been replaced by the quasineutrality condition. The dust fluid velocity  $\vec{v}_d$ is normalized by  ${\lambda_d}{\omega_{pd}}$. The term ${\tau_{m}}({\vec{v}_d}{\cdot}{\nabla})$ in the generalized momentum equation is responsible for introducing the collective behavior in the medium. When this term becomes zero ($\tau_m$=0), the momentum equation becomes Navier-Stokes equation. In other words, the GHD model turns into a standard hydrodynamic fluid model. Moreover, the presence of this term conserves the Gallilean invariance \cite{tiwari2012longitudinal}.

%Next, the GHD model is extended to include the passive tracer particles which are suppose to be dispersed throughout the VE fluid. 
In our earlier manuscript \cite{dharodi2014visco},  we proposed an idea of extending the GHD model by including the passive tracer particles in interest to estimate of the diffusion in a medium. In order to accomplish this, we consider two kinds of point-like tracer particles, (i) non-inertial tracers, and (ii) inertial tracers. The non-inertial tracers follow the flow exactly while the velocity of inertial tracers differs from flow velocity due to viscous drag force~\cite{sapsis2010clustering}. The tracer particles dynamics is simulated using a one-way coupled Lagrangian point-particle approach. One way coupled particle approach means the tracers do not affect the fluid motion, in other words the tracers are passive.  We also neglect any kind of interaction between particles and gravity effect on their dynamics. We further assume that the inertial tracer particles with density $\rho_p$ different than the density $\rho_d$ of the fluid. Under these assumptions, the particles are transported in the flow according to the equations:
%~~~~~~~~~~~~~~~~~~~~~~~~~~~~~~~~~~~~~~~
\begin{eqnarray}\label{eq:inertial1}
	\frac{d{\vec v_{pi}}}{dt} &= \frac{1}{\tau_s}({\vec v_d}({\vec r_{pi}})-{\vec v_{pi}}){,}
\end{eqnarray}
%~~~~~~~~~~~~~~~~~~~~~~~~~~~~~~~~~~~~~~~
\begin{eqnarray}\label{eq:inertial2}
	\frac{d{\vec r_{pi}}}{dt} &= {\vec v_{pi}}{,}
\end{eqnarray}
%~~~~~~~~~~~~~~~~~~~~~~~~~~~~~~~~~~~~~~~
where ${\vec r_{pi}}$ and $\vec{v}_{pi}$ are the position and velocity of the $i$th particle, respectively. ${\vec v_d}({\vec r_{pi}})$ is the dust fluid velocity at the particle position, ${\vec r_{pi}}$ which is obtained by solving the set of Eqs. (\ref{eq:vort_num}) and (\ref{eq:psi_num}). These equations are a simplified approximation of the Maxey-Riley equations \cite{maxey1983equation}.  Although, the neglected effects might have significant impacts in real flows, but these could be incorporated into the future study because even after neglecting them, the Eq. (\ref{eq:inertial1}) describe an enough complex system and it is worth studying to set the foundation for the future research. The particle time-scale $\tau_s={2{r^2_0}{\rho_p}}/{(9{\eta}{\rho_d})}$  denotes the response time of the particles is known as Stokes time \cite{guha2008transport}. $r_0$ is the radius of a particle. %$r_0=\sqrt[3]{3/{4{\pi}{\rho_p}}}$.

Although the particles are assumed as point particles, but they do have  finite mass and therefore finite inertia. The ratio of a particle time-scale to a fluid time-scale is known as Stokes number (St). The effect of particle inertia is often given by using the $St$ or $\tau_s$. On the basis of St or $\tau_s$ or inertia: For low value, the particles are predicted to follow the fluid flow passively like fluid particles, while at very high value almost the particles remain unaffected by the medium fluctuations. In-between these two limits,  when the particle and fluid time-scales are comparable, the particles respond in fast and strong manner  to the fluctuations.

The non-inertial tracers follow the fluid flow exactly and can be considered as attached to fluid surface. They are characterised by their position ${\vec r_{pi}}$, and velocity ${\vec v_{pi}} = {\vec v_d}({\vec r_{pi}})$ that is the dust fluid velocity at their position. Their equation of motion corresponds to the limit $\tau_s\rightarrow$ 0 in the set of Eqs. (\ref{eq:inertial1}) and (\ref{eq:inertial2}) becomes:
%~~~~~~~~~~~~~~~~~~~~~~~~~~~~~~~~~~~~~~~
\begin{eqnarray}\label{eq:non_inertial}
	\frac{d{\vec r_{pi}}}{dt}={\vec v_d}({\vec r_{pi}}){,}
\end{eqnarray}
%~~~~~~~~~~~~~~~~~~~~~~~~~~~~~~~~~~~~~~~
The set of Eqs. (\ref{eq:in_momentum})-(\ref{eq:in_continuity})-(\ref{eq:inertial1})-(\ref{eq:inertial2}) and Eqs. (\ref{eq:in_momentum})-(\ref{eq:in_continuity})-(\ref{eq:non_inertial})  represent the viscoelastic model for inertial and non-inertial tracer particles, respectively, in which the tracers follow the evolution of fluid with time in Lagrangian way. Both the set of equations would be referred as incompressible Generalized Hydrodynamic Tracer Transport (i-GHT2 or i-GHTT) model henceforth in the article. It should be noted that in i-GHTT model, in interest to  include both the compressible longitudinal and incompressible transverse modes just replace the i-GHD model ((set of Eqs. (\ref{eq:in_momentum})-(\ref{eq:in_continuity})) with complete GHD model (set of Eqs. (5)- (6)-(7) in \cite{kaw1998low}), say GHTT model .
   
  %%%%%%%%%%%%%%%%%%%%%%%%%%%%%%%%%%%%%%%
  % FIELD QUANTITIES FROM MD
  %%%%%%%%%%%%%%%%%%%%%%%%%%%%%%%%%%%%%%% 
  \subsection{Transport Properties}
  \label{}
  %~~~~~~~~~~~~~~~~~~~~~~~~~~~~~~~
  To quantify the average diffusion of tracer particles, the ensemble averaged mean square displacement of tracers is measured which is associated with the mixing of the fluid \cite{jeon2013anomalous}. The MSD is defined as,
%~~~~~~~~~~~~~~~~~~~~~~~~~~~~~~~~~~~~~~~
\begin{equation}\label{eq:MSD}
	MSD(t)= {\frac{1}{N}}{\sum^N_{j=1}}{(r_j(t)-r_j(0))^2}{,}
	\nonumber
\end{equation}  
%~~~~~~~~~~~~~~~~~~~~~~~~~~~~~~~
Here, $r_j(0)$ is initial position of $j$th particle at t=0 and $r_j(t)$ is position at time $t$. N is the total number of tracers in the ensemble. The slope of the MSD versus time is proportional to the diffusion coefficient of tracer particles which in turn is supposed to measure the mixing performance of the  carrier fluid. Thus, the mixing performance of the carrier fluid can be quantified through the time-MSD slope, the larger slope means the carrier fluid is a better mixture.

%%%%%%%%%%%%%%%%%%%%%%%%%%%%%%%%%%%%%
% Simulation methodology
%%%%%%%%%%%%%%%%%%%%%%%%%%%%%%%%%%%%% 
\section{Simulation methodology}
\label{simulation_metho}
%~~~~~~~~~~~~~~~~~~~~~~~~~~~~~~~
For the numerical simulations, first we need to express the model Eq. (\ref{eq:in_momentum}) as per requirements of simulation software, LCPFCT (Boris $et al.$ \cite{boris_book}). To fulfill these requirements split the Eq.~(\ref{eq:in_momentum}) in following two coupled equations, 
%~~~~~~~~~~~~~~~~~~~~~~~~~~~~~~~~~~~~~~~
\begin{eqnarray}\label{eq:vorticity4}
{{\frac{\partial \vec{v}_d } {\partial t}+\vec{v}_d  
		\cdot \nabla\vec{v}_d }+ \frac{\nabla P}{n_d} -\nabla \phi_d}={\vec  \psi} {,}
\end{eqnarray}
%~~~~~~~~~~~~~~~~~~~~~~~~~~~~~~~~~~~~~~~
\begin{eqnarray}\label{eq:psi4}
\frac{\partial {\vec \psi}} {\partial t}+\vec{v}_d \cdot \nabla{\vec \psi}=
{\frac{\eta}{\tau_m}}{\nabla^2}{\vec{v}_d }-{\frac{\vec \psi}{\tau_m}}{.}
\end{eqnarray}
%~~~~~~~~~~~~~~~~~~~~~~~~~~~~~~~~~~~~~~~
For  two-dimensional (2D) studies all the  variables are functions of  $x$ and $y$ only.  The new introduced quantity ${\vec \psi}(x,y)$ in LHS of Eq.~(\ref{eq:vorticity4}) represents the strain induced in the elastic medium by the time-varying velocity fields. Next, the gradient terms are eliminated by taking the curl of Eq.~(\ref{eq:vorticity4}) which yields an equation for the evolution of the vorticity field.  So the  coupled set of Eqs.~(\ref{eq:vorticity4})-(\ref{eq:psi4}) has been recast in the following form:
%~~~~~~~~~~~~~~~~~~~~~~~~~~~~~~~~~~~~~~~
\begin{equation}\label{eq:vort_num}
	\frac{\partial{\xi}_z} {\partial t}+\left(\vec{v}_d \cdot \vec \nabla\right)
	{{\xi}_z}={\vec
		\nabla}{\times}{\vec \psi}{,}  
\end{equation}
%~~~~~~~~~~~~~~~~~~~~~~~~~~~~~~~~~~~~~~~
\begin{eqnarray}\label{eq:psi_num}
\frac{\partial {\vec \psi}} {\partial t}+\vec{v}_d \cdot \nabla{\vec \psi}=
{\frac{\eta}{\tau_m}}{\nabla^2}{\vec{v}_d }-{\frac{\vec \psi}{\tau_m}}{.}
\end{eqnarray}
%~~~~~~~~~~~~~~~~~~~~~~~~~~~~~~~~~~~~~~~
Here, ${\xi_z}={\nabla}{\times}{\vec{v}_d}$ is the vorticity.  ${\vec \xi}$ is normalised with dust plasma frequency. The LCPFCT software \cite{boris_book} is based on a finite difference method has been used to solve the coupled set of Eqs. (\ref{eq:vort_num}) and (\ref{eq:psi_num}).  
Taking the curl of  relation  ${\xi_z}={\nabla}{\times}{\vec{v}_d}$ and using incompressible condition given by Eq. (\ref{eq:in_continuity}) i.e. $\nabla \cdot \vec{v}_d= 0$, we get
%~~~~~~~~~~~~~~~~~~~~~~~~~~~~~~~~~~~~~~~
\begin{eqnarray}\label{eq:vel-vort-rel}
{\nabla}^2{\vec{v}_d}  =-{\nabla}{\times}{\xi_z}
\end{eqnarray}
%~~~~~~~~~~~~~~~~~~~~~~~~~~~~~~~~~~~~~~~
This  velocity-vorticity relation is used to update the fluid velocity at each time step using FISHPACK \cite{swarztrauber1999fishpack}. The validation of our numerical code has been done in our earlier papers \cite{dharodi2014visco,dharodi2020rotating}.  

Further details on simulation procedure in advancing the tracer particles with flow of a VE fluid.  We have the dust fluid velocity for each particle at their respective position at each time step.  This dust velocity is going to be used in particle momentum Eq. (\ref{eq:inertial1}). Equations (\ref{eq:inertial1}) and  (\ref{eq:inertial2})  are numerically  integrated together to find the new position and velocity at the end of each time step.  This integration is based on the first order Runge-Kutta method. The particle velocity $\vec v_p$ is calculated by interpolating the velocity defined on nearby grid points, based on first-order Lagrangian interpolation scheme. The particles are advanced with fluid time step.

%%%%%%%%%%%%%%%%%%%%%%%%%%%%%%%%%%%%%
% Numerical results and discussion
%%%%%%%%%%%%%%%%%%%%%%%%%%%%%%%%%%%%% 
\section{Numerical results and discussion}
\label{results}
%~~~~~~~~~~~~~~~~~~~~~~~~~~~~~~~
%%%%%%%%%%%%%%%%%%%%%%%%%%%%%%%%%%%%%%%
The prime objectives of this section are to numerically explore the evolution of different types of sharp rotating vorticity flows in VE fluids and quantifying the mixing they generate using the tracer particles simulations. For each type of flow, the simulations are performed for the varying coupling strength of VE fluid which is usually measured as the ratio ${\eta/\tau_m}$. All the simulations are performed with periodic boundary conditions  in both the x and y directions on simulation box.

%Throughout the fluid simulations are performed with $1024{\times}1024$ grid points in both the x and y directions on simulation box with boundary periodic conditions.
%
%%%%%%%%%%%%%%%%%%%%%%%%%%%%%%%%%%%%%%%
%\section{Evolution of sharp vortex}
\subsection{KH instability of rotating SCDPs}
\label{KHevolution}
\begin{figure}%[ht]
	\centering   
	\includegraphics[width=0.5\linewidth]{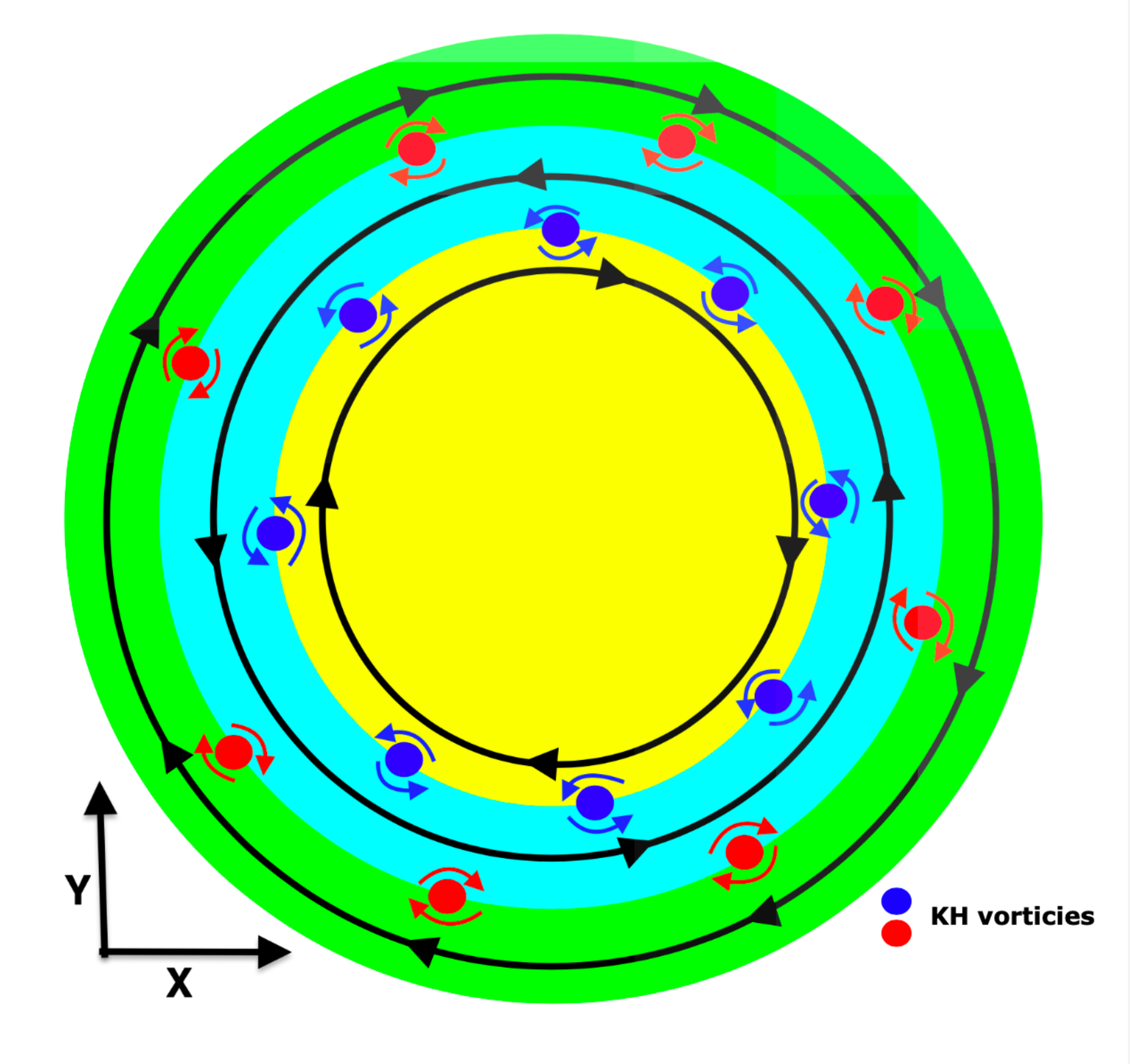}
	\caption{Formation of KH vortices at the circular sharp interfaces of core-shell flow regions (without any corresponding scale). One of flows forms the inner core and the others make the outer shells.  The black circles with arrows in the flow regions indicates the direction of rotation. At core-shell interface, the KH votices rotate anti-clockwise (blue color vortices with blue curved arrows) while at shell-shell interface rotate clockwise (red color vortices with red curved arrows).}
	\label{fig:kh_schematic}
\end{figure}

%~~~~~~~~~~~~~~~~~~~~~~~~~~~~~~~~~~~
Since our interest is in KH instability of rotating SCDPs, we consider the specific cases of sharp vorticity patches: (i) single-circulation, and (ii) multi-circulation vorticity shell profiles.  Before proceeding with the direct assessment of the evolution of KH instability through the numerical results,  it is good to understand the process of formation of these vortices for sharp rotating flows through the schematic picture illustrated in  Fig. \ref{fig:kh_schematic}. The multi-circulation vorticity profile can be depicted as core-shell fluid flows. In  Fig. \ref{fig:kh_schematic}, one of rotating flows forms the inner core (yellow color regime) and the others make the outer shells (cyan and green color regimes). Each rotating flow is divided by a sharp interface; the fluid on either side rotates in opposite directions. The black circle with arrows in the flow regime indicate its direction of rotation. 

%~~~~~~~~~~~~~~~~~~~~~~~~~~~~~~~~~~~
%\FloatBarrier
%~~~~~~~~~~~~~~~~~~~~~~~~~~~~~~~~~~~

At each of the interfaces, the counter-rotating flows create a region of high shear which immediately evolve into small co-rotating (like-sign) KH vortices. The schematic  diagram clearly illustrates how the  direction of rotation of these vortices depend on the relative motion between the alternative flows. At inner (core-shell) interface, the KH votices rotate anti-clockwise (blue color vortices with blue curved arrows) while at outer (shell-shell) interface rotate clockwise (red color vortices with red curved arrows). Thus far, these appraisals are particularly true for an inviscid HD fluid where no source term exists. Whereas an incompressible VE fluid, besides KH instability, would also support the TS waves that propagate through the medium at phase velocity $v_p= \sqrt{\eta/\tau_m}$.  A medium with higher coupling strength (ratio ${\eta/\tau_m}$) favors the faster and stronger TS waves. A stronger wave shows a less  fall in amplitude with time in comparison to the lower one (see Figs. 2 and 4 in \cite{dharodi2014visco}, and Fig. 4 in \cite{dharodi2020rotating}). Since, the present simulations have been carried out in the x-y plane, which is the plane of rotation of vorticity profile, the shear waves emitted from each interface should be cylindrical in shape. Thus, a stronger cylindrical wave tries to  dominate over KH instability and attempts to keep the cylindrical symmetry during the evolution.
 
In evolution of these KH vortices  the transport processes like merging and convection become important. When two co-rotating (like-sign) vortices are brought sufficiently close to each other they start to rotate around one another and eventually merge to form a single vortex while in convection the counter-rotating (unlike-sign) propagate together as single structure (dipole) to convect the fluid.

%%%%%%%%%%%%%%%%%%%%%%%%%%%%%%%%%%%%%
\subsubsection{Evolution of single-circulation sharp vorticity vortex}
\label{SubSec:sharp_single}
%~~~~~~~~~~~~~~~~~~~~~~~~~~~~~~~~~~~~~~~
The velocity profile for the single-circulation sharp vorticity vortex is given as follows
%~~~~~~~~~~~~~~~~~~~~~~~~~~~~~~~~~~~~~~~
\begin{equation}\label{eq:sim_sharp_vel}
{\vec v_{0}} = \left\{
\begin{array}{ll}
v_{x0}=-{\phi_0}{\frac{(y-y_c)}{b}};\: 
v_{y0}={\phi_0}{\frac{(x-x_c)}{a}}  & \quad |r| \leq 6 \\
0  & \quad \mathrm{otherwise}{.}
\end{array}
\right.
\end{equation}
%~~~~~~~~~~~~~~~~~~~~~~~~~~~~~~~~~~~~~~~
The  vorticity corresponding to the above velocity profile is given below
%~~~~~~~~~~~~~~~~~~~~~~~~~~~~~~~~~~~~~~~
\begin{equation}\label{eq:sim_sharp_vort}
{\xi_{z0}} = \left\{
\begin{array}{ll}
{\phi_0}{\left(\frac{1}{a}+\frac{1}{b}\right)} & \quad |r| \leq 6.0 
\\
0  & \quad \mathrm{otherwise}{.}
\end{array}
\right.
\end{equation}
%~~~~~~~~~~~~~~~~~~~~~~~~~~~~~~~~~~~~~~~
Here $|r|= {\sqrt{{\left({(x-x_c)/a}\right)}^2+{\left({(y-y_c)/b}\right)}^2}}$, $a$ and $b$ are the major and minor axes, respectively. $x_c$ and $y_c$ are the $x$ and $y$ coordinates of the center of the vorticity profile. The vorticity will have a clockwise rotation if amplitude ${\phi_0}{>}0$, or have a anti-clockwise rotation if ${\phi_0}{<}0$ and have no rotation if ${\phi_0}{=}0$.  

We consider the clockwise rotating vorticity profile has circular symmetry with parameters $a=b=1$,  amplitude ${\phi_0}=1$, sharp cutoff at radial distance $|r|$=6 units away from the centre of the vortex $(0,0)$. This vorticity profile (yellow color) is shown in the following Fig. \ref{fig:z_all_initial_vort}(a) which has circular interface with surrounding stagnant fluid  at $t=0$.  The simulation region is a square box of length 12$\pi$ units. 

%~~~~~~~~~~~~~~~~~~~~~~~~~~~~~~~~~~~
\begin{figure}
	\centering   
	\includegraphics[width=1\linewidth]{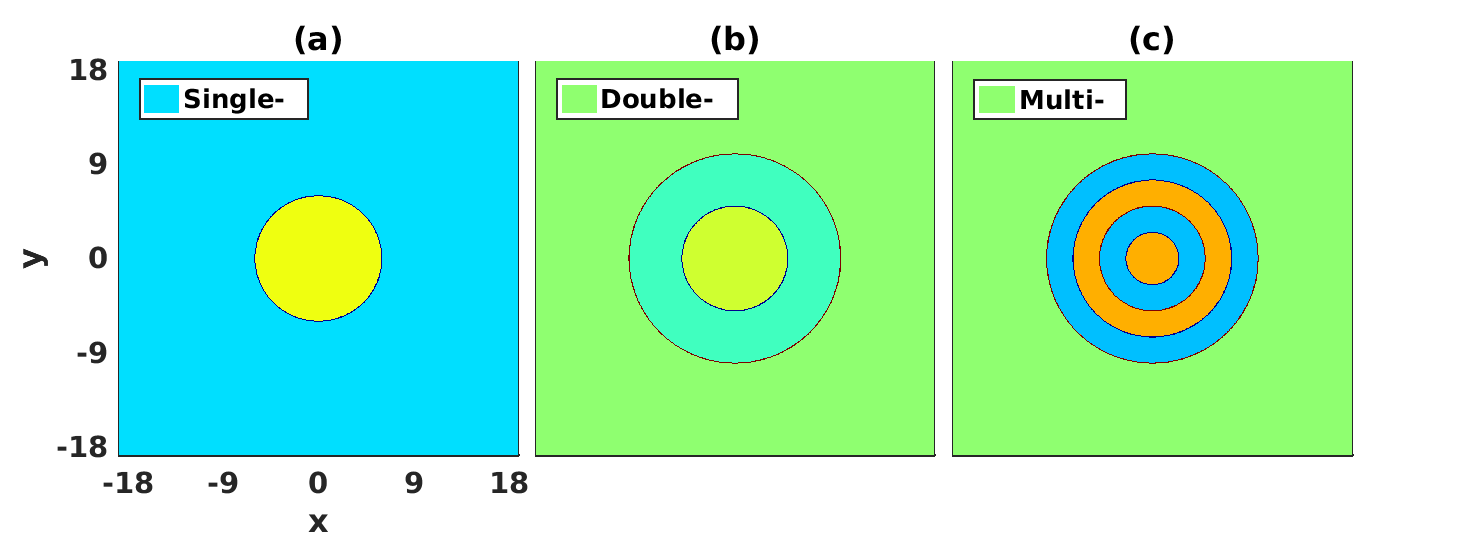}
	\caption {Initial rotating sharp vorticity profiles at t = 0 to study the KH instability. The color scale corresponding to a particular vorticity profile has been given in the following respective figure which shows its evolution. The simulation region is a square box of length 12$\pi$ units for all these systems.
	}
	\label{fig:z_all_initial_vort}
\end{figure}
%\FloatBarrier
%~~~~~~~~~~~~~~~~~~~~~~~~~~~~~~~~~~~

\begin{figure}
	\centering   
	\includegraphics[width=1.0\linewidth]{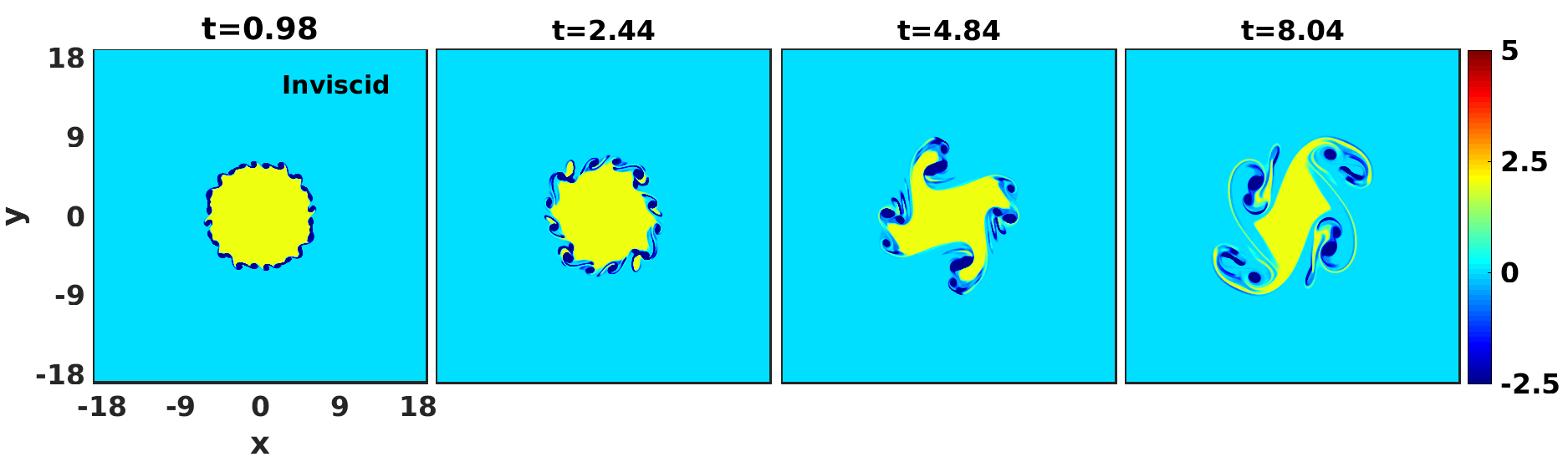}
	\caption {An inviscid HD fluid. The time evolution of a sharp clockwise rotating vorticity which has circular interface with surrounding stagnant fluid. The sharpness of the vorticity profile generates small KH vortices at the interface that results in a anisotropic isolated structure.}
	\label{fig:sharpfluidphi1}
\end{figure}

The panels of Fig. \ref{fig:sharpfluidphi1}  shows time evolution of vorticity profile given in Fig. \ref{fig:z_all_initial_vort}(a) in the inviscid HD fluid. The sharpness of the  clockwise rotating vorticity profile generates a  strong rotational sheared flow. This  sheared flow results in creation of small co-rotating (anti–clockwise) KH vortices (dark blue color vortices) at the vorticity interface $|r|=6$.  These like-sign vortices start merging as rotation progresses that leads the fluid to evolve into an anisotropic isolated structure.

%~~~~~~~~~~~~~~~~~~~~~~~~~~~~~~~~~~~
%\FloatBarrier
%~~~~~~~~~~~~~~~~~~~~~~~~~~~~~~~~~~~

\begin{figure}
	\centering                        
	\includegraphics[width=1\linewidth]{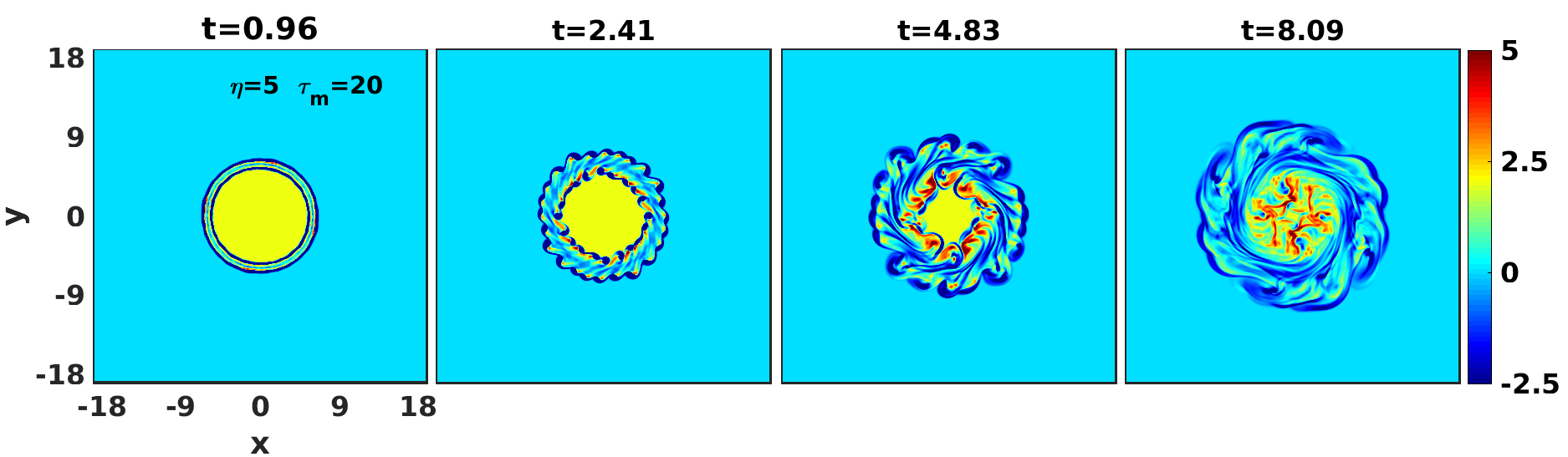}
		\caption {Viscoelastic fluid with coupling parameters ${\eta}=5$, and ${\tau_m}=20$. The time evolution of a sharp rotating vorticity which has circular interface with surrounding stagnant fluid. The sharpness of the vorticity profile generates small KH vortices at the interface along with a pair of ingoing and outgoing wavefronts of TS waves that assist in fluid mixing by convecting it inside and outside the vortex structure. }
	\label{fig:sharp520phi1}
\end{figure}

 In Fig. \ref{fig:sharp520phi1} where ${\eta}=5$ and ${\tau_m}=20$, once the vortex begins to rotate, we observe a pair of ingoing and outgoing  cylindrical shear waves from the interface along with these small like-sign KH vortices. During the evolution, it is observe that both the waves carry the like-sign vortices which interact with themselves in order to merge and have interplay with these waves as well. The fluid within the inner region  ($|r| \leq 6$) favors mixing due to the ingoing waves, while the stationary fluid in the outer region ($|r| \geq 6$) gets mixing due to the outgoing waves. Thus, the TS waves assist the process of fluid mixing by convecting it inside and outside the vortex structure. Unlike the inviscid case, here, the KH vortices are confined to the radial emitted waves.  Since the wavefronts are cylindrical in shapes, the interplay between waves and vortices  leads the evolution of the VE fluid toward an isotropic structure. 
 
%~~~~~~~~~~~~~~~~~~~~~~~~~~~~~~~~~~~~~~~
%~~~~~~~~~~~~~~~~~~~~~~~~~~~~~~~~~~~~~~~

\begin{figure}
	\centering               
	\includegraphics[width=1\linewidth]{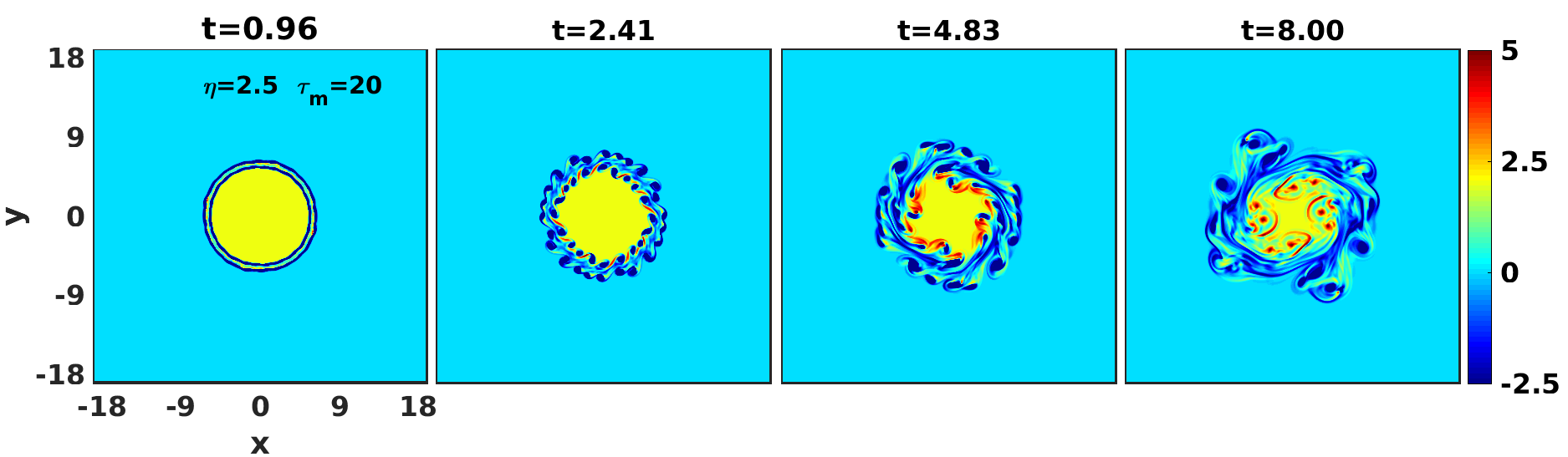}
		\caption {The time evolution of a sharp rotating vorticity in VE fluid with coupling parameters ${\eta}=2.5$, and ${\tau_m}=20$. Due to the less coupling strength, the evolution of medium towards an isotropic structure and mixing process are much slower in comparison to Fig. \ref{fig:sharp520phi1}.  
 }
	\label{fig:sharp2p520phi1}
\end{figure}

In Fig. \ref{fig:sharp2p520phi1}, we have simulated another case of VE fluid which has less coupling strength ($\eta/\tau_m$=0.125) with ${\eta}=2.5$ and ${\tau_m}$=20. Unlike Fig. \ref{fig:sharp520phi1}, since the TS waves are not strong enough that they can dominant over the KH instability, less confinement of KH vortices to the waves. This is evident from Fig. \ref{fig:sharp2p520phi1}  where the evolution of medium towards an isotropic structure and mixing process are much slower in comparison to Fig. \ref{fig:sharp520phi1}.  

%~~~~~~~~~~~~~~~~~~~~~~~~~~~~~~~~~~~~~~~
%\FloatBarrier
%~~~~~~~~~~~~~~~~~~~~~~~~~~~~~~~~~~~~~~~

 In conclusion, as a result of greater the coupling strength or stronger TS wave, the medium evolution attempts to possess radial symmetry and shows better mixing. The mixing is found minimal in inviscid fluid.

The single-circulation sharp profile carries single types of anti-clockwise  rotating KH vortices across its only interface. While a multi-circulation profile with two or more than two interfaces produces  clockwise as well as anti-clockwise KH vortices. Thus, in multi-circulation case apart from merging between like-sign vortices, the propagation of unlike-sign vortices  as a dipolar structure also becomes important which can assist to increasing the spatial domain of mixing fluids.

%%%%%%%%%%%%%%%%%%%%%%%%%%%%%%%%%%%%%%%%%%
\subsubsection{Multi-circulation vorticity shell profile}
\label{difussionch3}
\begin{figure}
	\centering         
	\includegraphics[width=1\linewidth]{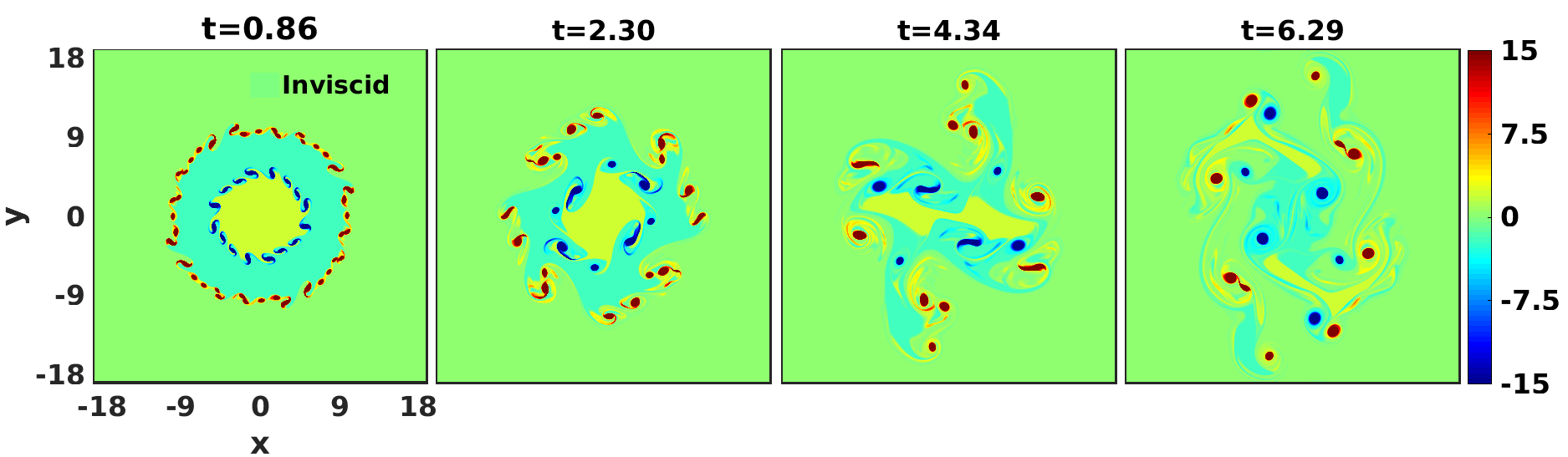}
		\caption {An inviscid HD fluid. Evolution of of vorticity having two circular sharp flows, inner core and an outer shell flows have reversal circulation, in time generates small KH vortices at each interface that results in a anisotropic isolated structure.}
	\label{fig:doublesharpfluidphi1}	 
\end{figure}

%~~~~~~~~~~~~~~~~~~~~~~~~~~~~~~~~~~~~~~~ 
Here, we first consider the simplest case of multiple shells of vorticity having two flows: Inner core flow and a outer shell flow, both flows have reversal circulation. The velocity profile for this configuration is given below.  
%~~~~~~~~~~~~~~~~~~~~~
\begin{equation}\label{eq:double_vel}
	{\vec v_{0}} = \left\{
	\begin{array}{ll}
		v_{x0}=-{\phi_0}{(y-y_c)};\: v_{y0}={\phi_0}{(x-x_c)}  & \quad |r| 
		\leq 5 \\
		v_{x0}={\phi_0}{(y-y_c)};\: v_{y0}=-{\phi_0}{(x-x_c)}  & \quad 
		5<|r|\leq 10 \\
		0  & \quad \mathrm{otherwise}{.}
	\end{array}
	\right.
\end{equation}
%~~~~~~~~~~~~~~~~~~~~~~~~~~~~~~~~~~~
The vorticity corresponding to the above velocity profile is given below,
%~~~~~~~~~~~~~~~~~~~~~~~~~~~~~~~~~~~
\begin{equation}\label{eq:double_vort}
	{\xi_{z0}} = \left\{
	\begin{array}{ll}
		2{\phi_0} & \quad |r| \leq 5 \\
		-2{\phi_0} & \quad 5 < |r| \leq 10 \\
		0  & \quad \mathrm{otherwise}{.}
	\end{array}
	\right.
\end{equation}
%~~~~~~~~~~~~~~~~~~~~~~~~~~~~~~~~~~
With the parameter $\phi_0$=1, we consider both flows rotate with equal rotation rates and in opposite directions (Fig. \ref{fig:z_all_initial_vort}(b)). Figure \ref{fig:doublesharpfluidphi1} shows the evolution of this vorticity profile for an inviscid fluid.  At inner interface, the KH vortices rotate anti-clockwise (blue color vortices) while at outer interface rotate clockwise (red color vortices). As time goes on, the merging between like-sign vortices take place at both the interfaces, and simultaneously growing closeness between both the interfaces due to the radial gradient in vorticity results in interactions between the counter-rotating (red-blue) vortices as well.  These counter-rotating vortices results in the formation of propagating dipolar structures which help to convect the fluid across the wider domain than the single interface (Fig. \ref{fig:sharpfluidphi1}) which only favors the merging process.

%~~~~~~~~~~~~~~~~~~~~~~~~~~~~~~~~~~~~~~~~~~
%\FloatBarrier
%~~~~~~~~~~~~~~~~~~~~~~~~~~~~~~~~~~~~~~~

\begin{figure}
	\centering   	 
	\includegraphics[width=1\linewidth]{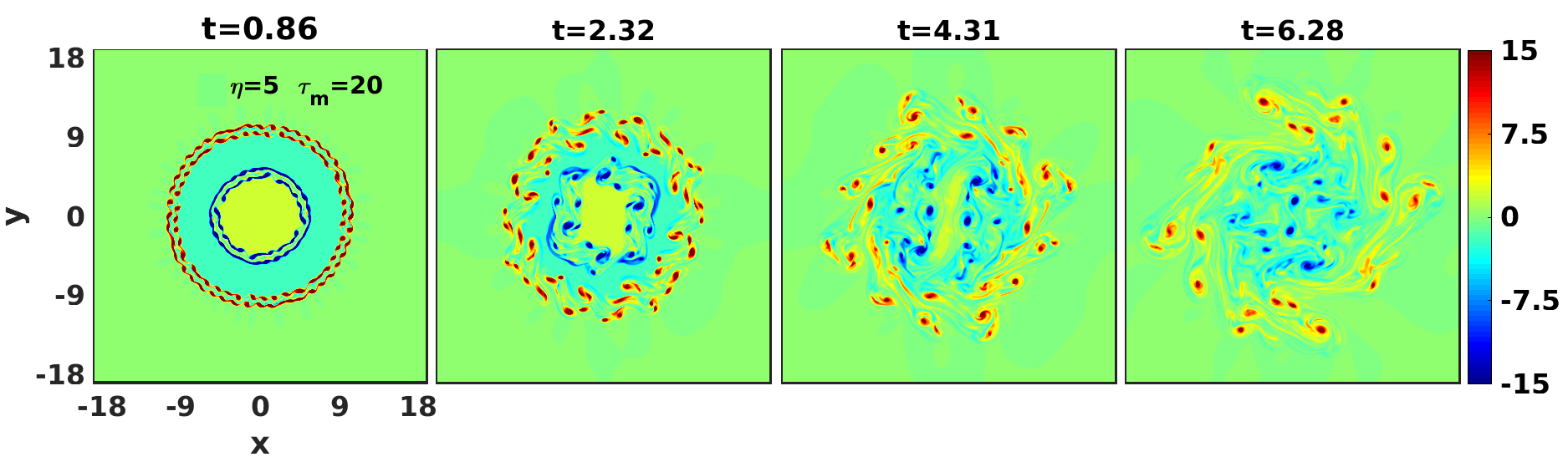}
	\caption {Viscoelastic fluid with coupling parameters ${\eta}=5$, and ${\tau_m}=20$. The time evolution of vorticity having two circular sharp interfaces, inner core and outer shell have reversal circulation, generates small KH vortices at each interface along with a pair of ingoing and outgoing  TS waves fronts that results in well fluid mixing by convecting it inside and outside the vortex structure.}
	\label{fig:doublesharp520phi1}
\end{figure}

In Fig.~\ref{fig:doublesharp520phi1}, we observe that a pair of ingoing and outgoing wavefronts emanates from each of the two sharp interfaces of the vortex structure along with KH vortices (see the first panel).  It is observe that both the wavefronts at inner interface carry the blue color (rotating anti-clockwise) vortices while at outer interface carry red color (rotating clockwise) vortices. During the emission of these shear waves like-sign vortices interact with themselves in order to get merge.

%~~~~~~~~~~~~~~~~~~~~~~~~~~~~~~~~~~~~~~~~~~
%\FloatBarrier
%~~~~~~~~~~~~~~~~~~~~~~~~~~~~~~~~~~~~~~~~~~~~~~

\begin{figure}
	\centering         
	\includegraphics[width=1\linewidth]{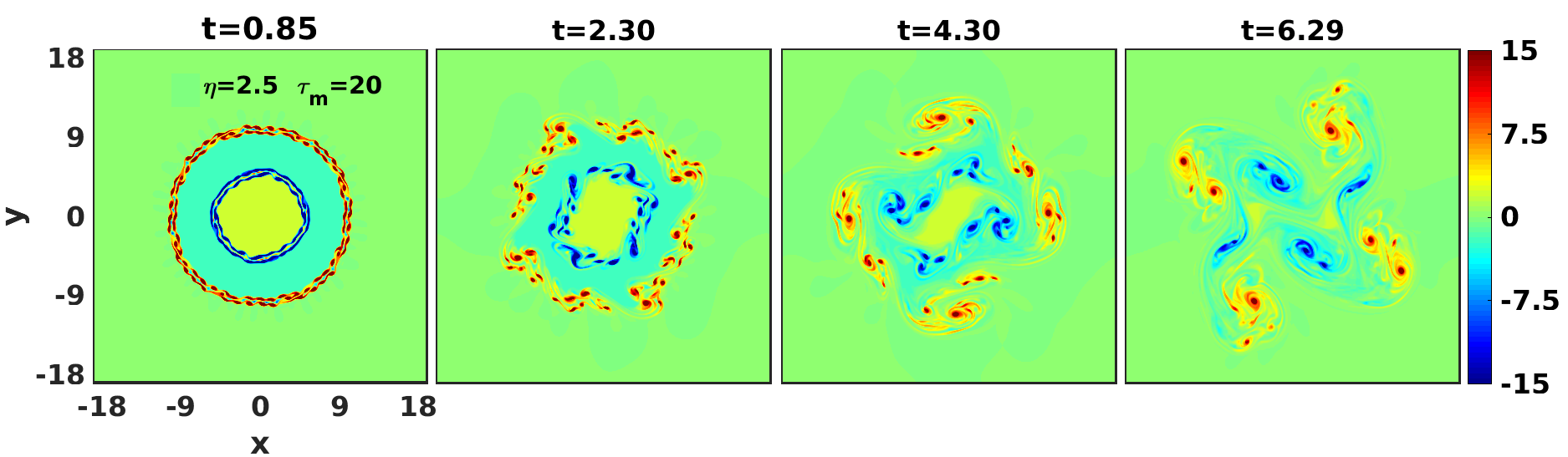}
	\caption {Viscoelastic fluid with coupling parameters ${\eta}=2.5$, and ${\tau_m}=20$. The time evolution of vorticity having two circular sharp interfaces, both have reversal circulation. Due to the less coupling strength the spatial confinement of this medium is lower than Fig.~\ref{fig:doublesharp520phi1}.
	}
	\label{fig:double2p520phi1}	 
\end{figure}

  The stagnant fluid in the outermost region ($|r| \ge 10$ )  undergoes mixing due to the outgoing wave from the outermost interface at 10, while the innermost vortex region undergoes mixing due to the ingoing wave emanating  from the sharp interface located at 5. Interestingly, the vortex region confined within the two sharp interfaces ($5 <|r|\leq 10$) undergoes mixing due to the ingoing wave from the outermost interface and the outgoing wave from the innermost interface. This region is probably convection dominating region due to the higher possibility of formation of propagating dipolar structures.   As the results of multiple interaction processes the mixing becomes fast and efficient than the cases discussed so far.   Next, we consider the another VE fluid in Fig.~\ref{fig:double2p520phi1} having lower coupling strength i.e. ${\eta}=2.5, {\tau_m}=20, v_p$=0.35, it is observed that at each time step in the spatial confinement of this medium is lower than Fig.~\ref{fig:doublesharp520phi1}. Moreover, the comparison manifests that the mixing and evolution symmetry of a medium are proportional to the coupling strength of that medium as observed for earlier cases.
  
%~~~~~~~~~~~~~~~~~~~~~~~~~~~~~~~~~~~~~~~~~~
%\FloatBarrier
%~~~~~~~~~~~~~~~~~~~~~~~~~~~~~~~~~~~~~~~

\begin{figure}
	\centering
	\includegraphics[width=1.0\textwidth]{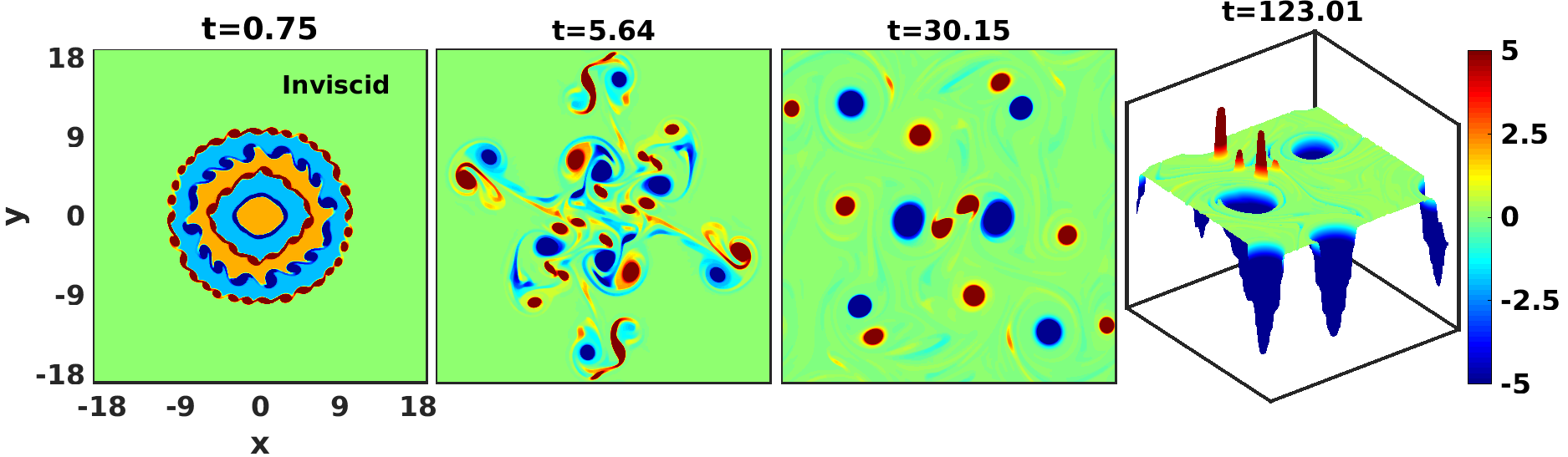}
	\caption {An inviscid HD fluid. The time evolution of  multi-circulation vorticity profile, each consecutive one having a reversal in its circulation, generates small KH vortices at each interface. This evolution provides a complete picture of a turbulent flow which is collection of various kind of vortices, and exhibits transport properties like convection and merging.}
	\label{fig:mixfliuid}
\end{figure}

Next, in order to make a better comparative numerical analysis about the mixing rate,  we have considered a more complex scenario of multiple circulations (each consecutive one having a reversal in its circulation) having the following velocity flow profile: 
%~~~~~~~~~~~~~~~~~~~~~
\begin{equation}
	{\vec v_{0}}= \left\{
	\begin{array}{ll}
		v_{x0}=-{\phi_0}{(y-y_c)};\: v_{y0}={\phi_0}{(x-x_c)}  & \quad |r| 
		\leq 2.5 \\
		v_{x0}={\phi_0}{(y-y_c)};\: v_{y0}=-{\phi_0}{(x-x_c)}  & \quad 
		2.5<|r|\leq 5 \\
		v_{x0}=-{\phi_0}{(y-y_c)};\: v_{y0}={\phi_0}{(x-x_c)}  & \quad 
		5<|r|\leq 7.5 \\
		v_{x0}={\phi_0}{(y-y_c)};\: v_{y0}=-{\phi_0}{(x-x_c)}  & \quad 
		7.5<|r|\leq 10 \\
		0  & \quad \mathrm{otherwise}{.}
	\end{array}
	\right.
\end{equation}
%~~~~~~~~~~~~~~~~~~~~~~~~~~~~~~~~~~~
The vorticity corresponding to the above velocity profile is given below,
%~~~~~~~~~~~~~~~~~~~~~~~~~~~~~~~~~~~
\begin{equation}
	{\xi_{z0}} = \left\{
	\begin{array}{ll}
		2{\phi_0} & \quad |r| \leq 2.5 \\
		-2{\phi_0} & \quad 2.5 < |r| \leq 5 \\
		2{\phi_0} & \quad 5 < |r| \leq 7.5 \\
		-2{\phi_0} & \quad 5 < |r| \leq 10 \\
		0  & \quad \mathrm{otherwise}{.}
	\end{array}
	\right.
\end{equation}
%~~~~~~~~~~~~~~~~~~~~~~~~~~~~~~~~~~
For the parameter $\phi_0$=1, the initial vorticity  profile is shown in Fig. \ref{fig:z_all_initial_vort}(c). The complexity of this motion of multi-circulation structure is evident from the subplots of Fig.~\ref{fig:mixfliuid} for inviscid fluid. In initial time period, the vortices of KH instability develop across the interface of each shell.  At intermediate time range, this evolution provides a complete picture of a turbulent flow throughout the entire system which is collection of several small symmetric and non-symmetric vortices. The transport phenomena like merging between two co-rotating vortices, convection due to the evolution of dipolar (two counter-rotating ) vortices, collision between these dipolar vortices, and also the formation and evolution of triplor structures becomes more frequent than above discussed cases.  

%~~~~~~~~~~~~~~~~~~~~~~~~~~~~~~~~~~~~~~
%\FloatBarrier
%~~~~~~~~~~~~~~~~~~~~~~~~~~~~~~~~~~~~~~~~

\begin{figure}
	\centering
	\includegraphics[width=1.0\textwidth]{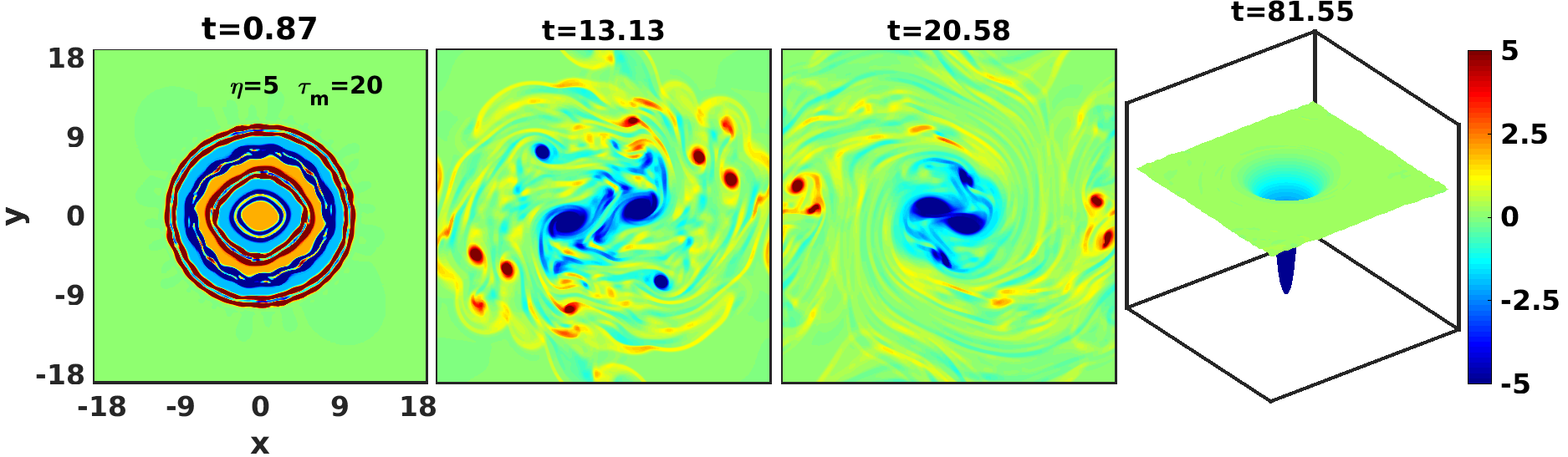}
	\caption {The time evolution of  multi-circulation vorticity profile in VE fluid with ${\eta}=5$, and ${\tau_m}=20$. The strong interaction between  KH vortices and TS waves results in the relaxing the medium into a single vortex faster than inviscid fluid.
	}
	\label{fig:mixghd520}
\end{figure}

Figure~\ref{fig:mixghd520}, represents the evolution of same initial profile (Fig.~\ref{fig:mixfliuid}) of vorticity for VE fluid with coupling parameters ${\eta}=5$, ${\tau_m}$=20.  From the comparative observations between Fig.~\ref{fig:mixfliuid} and Fig.~\ref{fig:mixghd520}, it is interesting to notice that the presence of TS waves leads to the relaxing of the turbulent medium to a single vortex faster than in inviscid fluid.

%~~~~~~~~~~~~~~~~~~~~~~~~~~~~~~~~~~~~~~~~~~~~~~~~~~~~~~~
%\FloatBarrier
%~~~~~~~~~~~~~~~~~~~~~~~~~~~~~~~~~~~~~~~~~~~~~~~~~~~~~~~~

\begin{figure}
	\centering
	\includegraphics[width=1.0\textwidth]{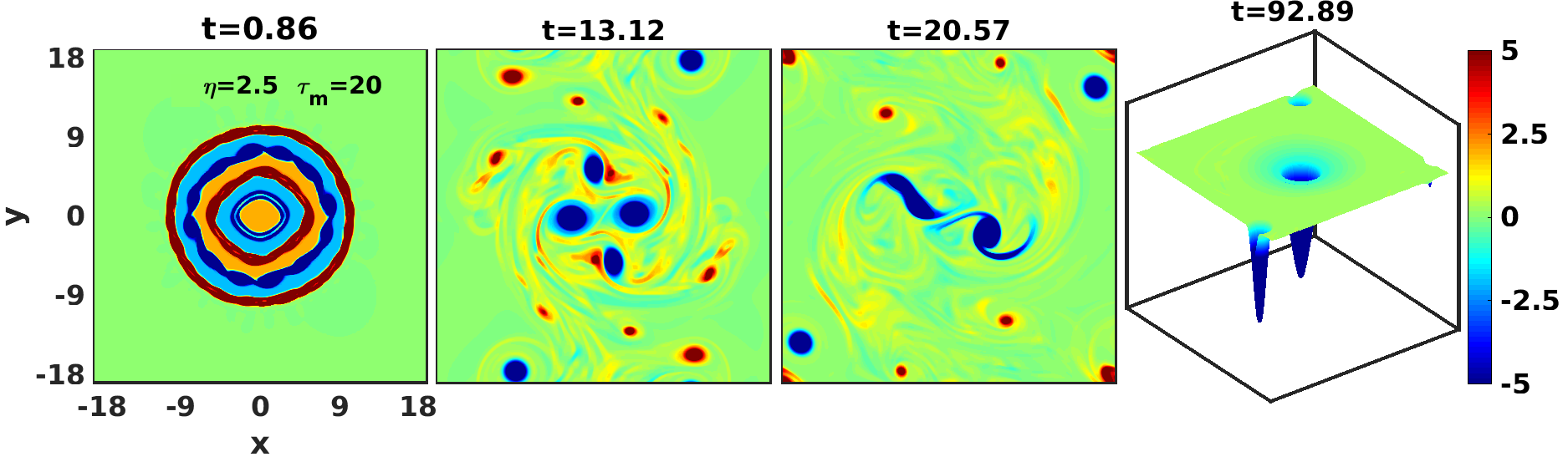}
	\caption { The time evolution of  multi-circulation vorticity profile in VE fluid with ${\eta}=2.5$, and ${\tau_m}=20$. Since the TS waves are weaker than Fig.~\ref{fig:mixghd520}, the relaxation time of the medium becomes longer.
}
	\label{fig:mixghd2p520}
\end{figure}

In Fig.~\ref{fig:mixghd2p520} (${\eta}=2.5, {\tau_m}=20, v_p$=0.35), since the TS waves are weaker than 
Fig.~\ref{fig:mixghd520}, the less spatial confinement of the turbulent medium that results in slower merging process, due to which the relaxation time of the medium becomes longer.

%%~~~~~~~~~~~~~~~~~~~~~~~~~~~~~~~~~~~~~~~~~~~~~~~~~~~~~~~
%\FloatBarrier
%~~~~~~~~~~~~~~~~~~~~~~~~~~~~~~~~

 The comparison of Fig.~\ref{fig:mixghd520} and Fig.~\ref{fig:mixghd2p520} clearly displays these observations.

%%%%%%%%%%%%%%%%%%%%%%%%%%%%%%%%%
%Advection of passive tracer particles
% Diffusion and clustering of inertial and non-inertial test particles
%%%%%%%%%%%%%%%%%%%%%%%%%%%%%%%%
\subsection{Transport of tracer particles}
\label{Clustering_chp4}
\begin{figure}
	\centering
	\includegraphics[width=1.0\textwidth]{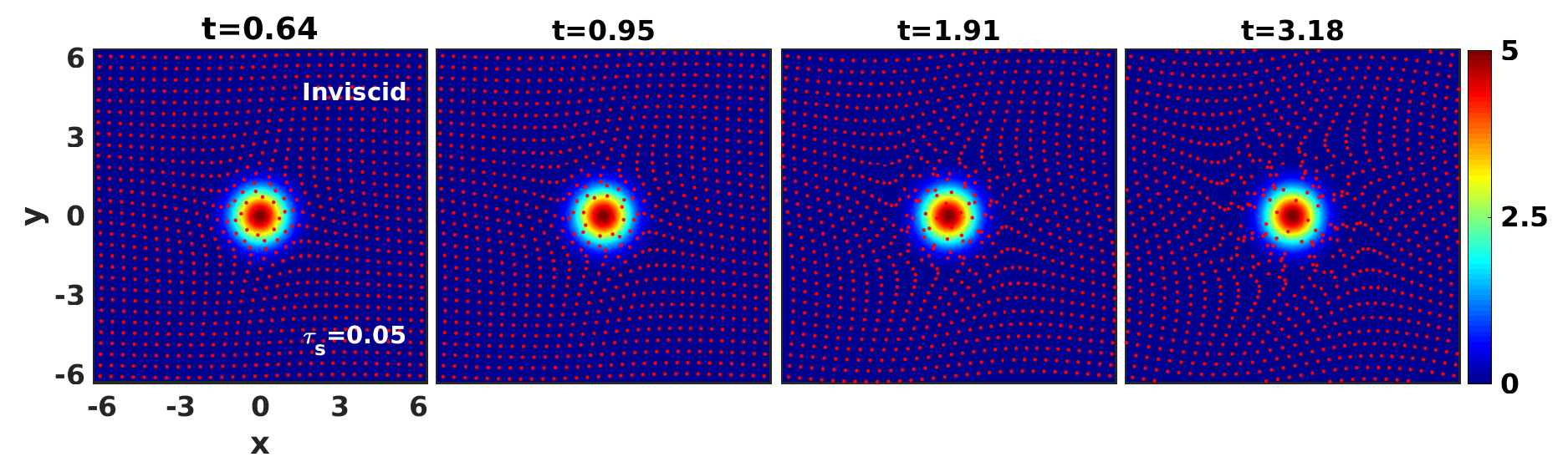}
		\caption {An inviscid HD fluid carries tracer particles are shown as red dots.  The low inertial tracers ($\tau_s$=0.05) follow the dynamics along the smooth rotating vortex.
		}
	\label{fig:particle_fluid_lighter}
\end{figure}

\begin{figure}
	\centering
	\includegraphics[width=1.0\textwidth]{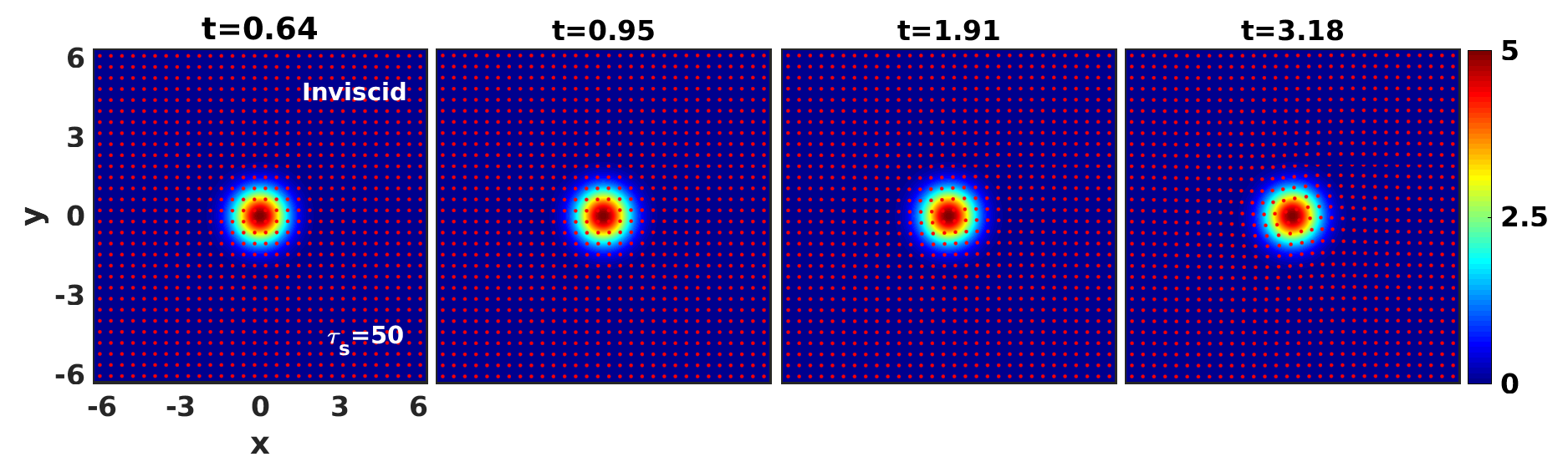}
		\caption {An inviscid HD fluid carries tracer particles are shown as red dots. The high inertial tracers ($\tau_s$=50) show negligible response to the rotating vortex.}
	\label{fig:particle_fluid_heavy}
\end{figure}

\begin{figure}
	\centering
	\includegraphics[width=1.0\textwidth]{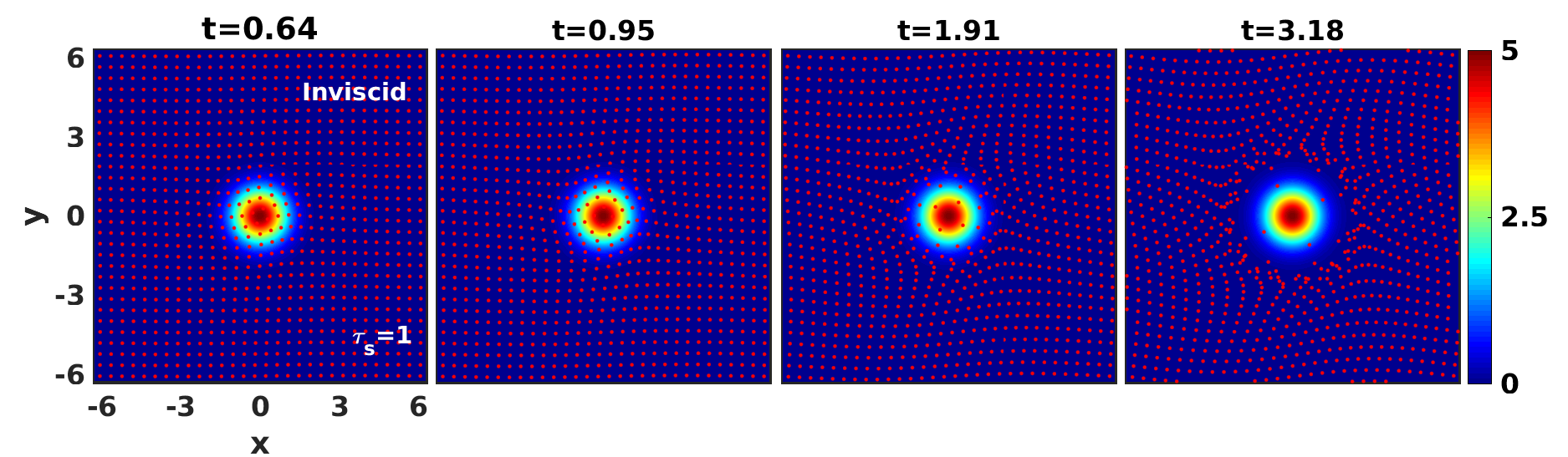}
		\caption {An inviscid HD fluid carries tracer particles are shown by red dots. Smooth rotating vorticity vortex  with tracers having intermediate inertia i.e $\tau_s$=1 show high response to the vorticity gradient, the particles are pushed away where the flow is strong enough and get accumulate in strain-dominated region.}
	\label{fig:particle_fluid_moderate}
\end{figure}

%~~~~~~~~~~~~~~~~~~~~~~~~~~~~~~~~~~~~~~~
In Section~\ref{GHT2} we have stated that tracer particles with very low inertia  follow the flow passively, while particles with very high inertia will remain almost unaffected by the medium fluctuations. In between these two limits particles show the strongest response to the medium fluctuations. The simulations are performed for all these three inertial particles: very low ($\tau_s$=0.05),  intermediate ($\tau_s$=1), and very high inertia ($\tau_s$=50).  To observe the exclusive effect of inertia, we transport these particles through a similar smooth rotating vorticity vortex in an inviscid fluid.  An inviscid fluid has no source term which favors the emission of TS waves or dissipative term like viscosity.  And also, we choose the smooth rotating vorticity vortex which does not satisfy the KH destabilization condition anywhere in the vorticity patch. The equation of such smooth rotating vorticity is given as 
%~~~~~~~~~~~~~~~~~~~~~~~~~~~~~~~~~~~~~~~
\begin{equation}\label{eq:smooth_vort}
	{\xi_{0}}={\Omega_0}exp\left(-\left({x^2+y^2}\right)/{a^2_c}\right){.}
\end{equation}
%~~~~~~~~~~~~~~~~~~~~~~~~~~~~~~~~~~~~~~~
The evolution of same structure given by Eq. \ref{eq:smooth_vort} for ${a_c}$=1.0,  ${\Omega_0}=5$ in an inviscid fluid is shown in Figs.~(\ref{fig:particle_fluid_lighter}),~(\ref{fig:particle_fluid_heavy}),  and (\ref{fig:particle_fluid_moderate}). From these figures it is clear that the rotating vortex keeps rotating without any change. 

%~~~~~~~~~~~~~~~~~~~~~~~~~~~~~~~~~~~~~~~
%~~~~~~~~~~~~~~~~~~~~~~~~~~~~~~~~~~~~~~~

Now, in order to see the response of tracer particles, initially  $(t=0)$, we have distributed 900  inertial particles (shown by red dots) homogeneously throughout the domain.

%~~~~~~~~~~~~~~~~~~~~~~~~~~~~~~~~~~~~~~~
%~~~~~~~~~~~~~~~~~~~~~~~~~~~~~~~~~~~~~~~

From Fig.~\ref{fig:particle_fluid_lighter}, it is clear that low inertial particles ($\tau_s$ =0.05) follow the dynamics along the rotating vortex, and the particles with higher inertia $\tau_s$= 50 (Fig.~\ref{fig:particle_fluid_heavy}) show negligible response to the vorticity gradient.

%~~~~~~~~~~~~~~~~~~~~~~~~~~~~~~~~~~~~~~~
%\FloatBarrier
%~~~~~~~~~~~~~~~~~~~~~~~~~~~~~~~~~~~~~~~

In comparison to previous cases (Fig.~\ref{fig:particle_fluid_lighter} and Fig.~\ref{fig:particle_fluid_heavy}), Fig.~\ref{fig:particle_fluid_moderate} shows that the particles with intermediate value of $\tau_s$=1 counter a significant outward push. It is because the particle and fluid time- scales are comparable which results the particles experience a notable centrifugal force due to vorticity gradient. Since, the  inertial particles are pushed away from regions where the flow is strong enough, these particles get accumulate in strain-dominated regions. Thus, particles tend to leave regions of high vorticity and cluster into regions of high strain \cite{ravichandran2017clustering}. Note, we have simulated a range of intermediate inertial particles, and observed the same effect with varying outward push.

With the identification of intermediate inertial particles and  understanding of their evolution, next we compare the evolution of intermediate inertial (typical, $\tau_s$=1) with non-inertial particles for the inviscid HD and VE ($\eta=5$, $\tau_m=20$)  fluids in the following Fig. \ref{fig:passive_particle_sharp_fluid} and Fig.~\ref{fig:st1_particle_sharp}, respectively. For this, we choose a sharp rotating vorticity profile which is given by Eq. (\ref{eq:sim_sharp_vort}). The middle row in Fig.~\ref{fig:passive_particle_sharp_fluid} and Fig.~\ref{fig:st1_particle_sharp} represent the time evolution of this profile for inviscid and VE fluids, respectively. As the sharp vortex starts to rotate, produces larger strain (deformation) in medium along the interface that results in formation of KH instability. The evolution of this rotor has already been discussed in the earlier Section~\ref{KHevolution} for inviscid fluid in Fig.~\ref{fig:sharpfluidphi1} and for VE fluid in Fig.~\ref{fig:sharp520phi1}. Initially $(t=0)$, here, we distribute 3600 tracer particles homogeneously throughout these fluids.  

%~~~~~~~~~~~~~~~~~~~~~~~~~~~~~~~~~~~~~~~
\begin{figure}
	\centering
	\includegraphics[width=1.0\textwidth]{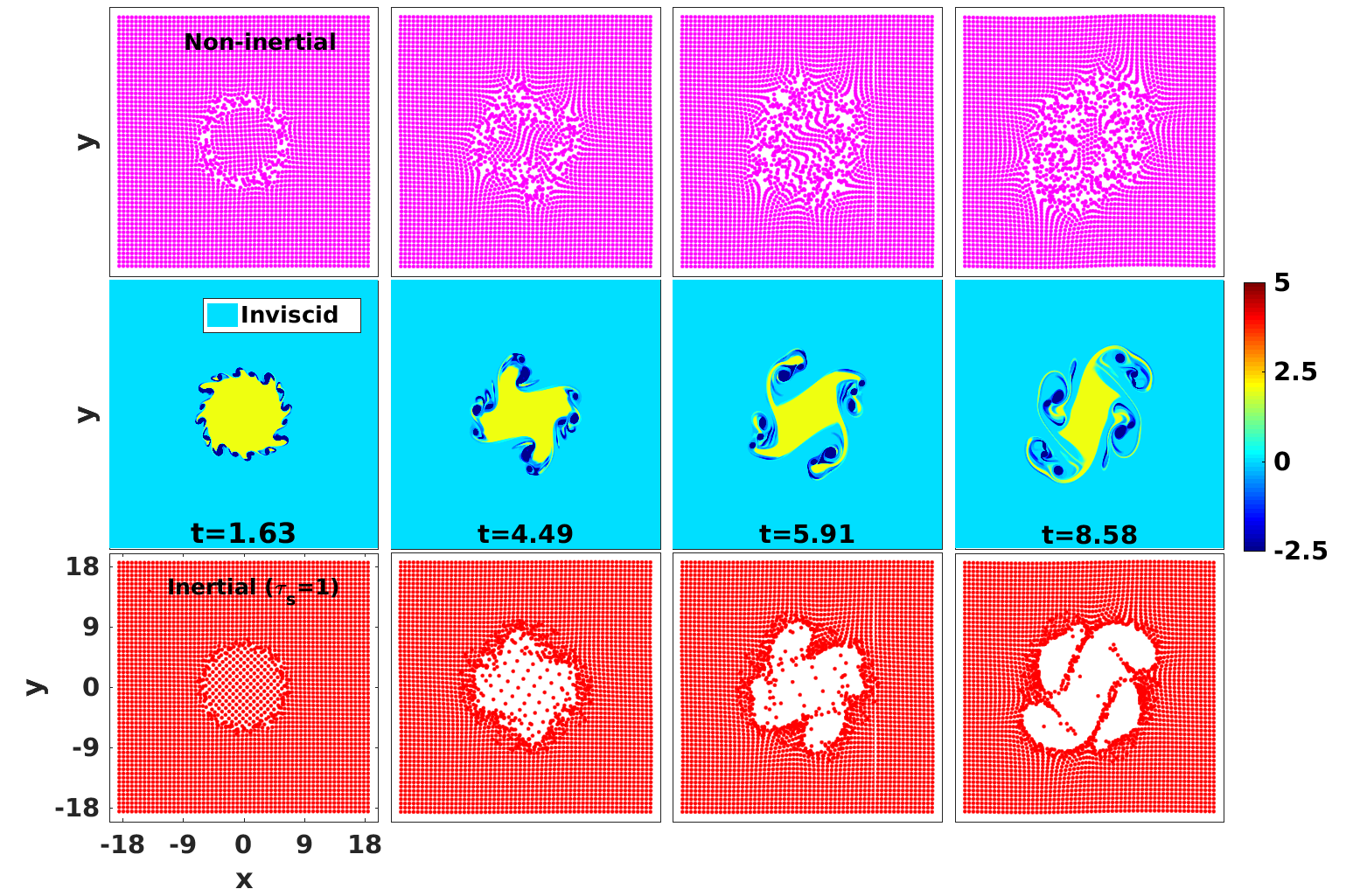}	
	\caption {An inviscid HD fluid advect tracer particles. First and third row show the temporal and spatial distribution of non-inertial particles (shown by magenta dots) and inertial particles ($\tau_s$=1, shown by red dots), respectively, corresponding to the sharp vorticity profile evolution shown in second row.} 
	\label{fig:passive_particle_sharp_fluid}
\end{figure}
%\FloatBarrier
%~~~~~~~~~~~~~~~~~~~~~~~~~~~~~~~~~~~~~~~

In both Figs. \ref{fig:passive_particle_sharp_fluid} and ~\ref{fig:st1_particle_sharp}, the first and the third row visualize the distribution of non-inertial particles (shown by magenta dots) and inertial particles (shown by red dots) respectively, advect by their  respective fluid flow shown in the middle row. The accumulation of non-inertial particles is observed in rotation-dominated regions over the vortex structures. While the inertial particles accumulate in strain-dominated regions along the interfaces.

%~~~~~~~~~~~~~~~~~~~~~~~~~~~~~~~~~~~~~~~
\begin{figure}
	\centering
	\includegraphics[width=1.0\textwidth]{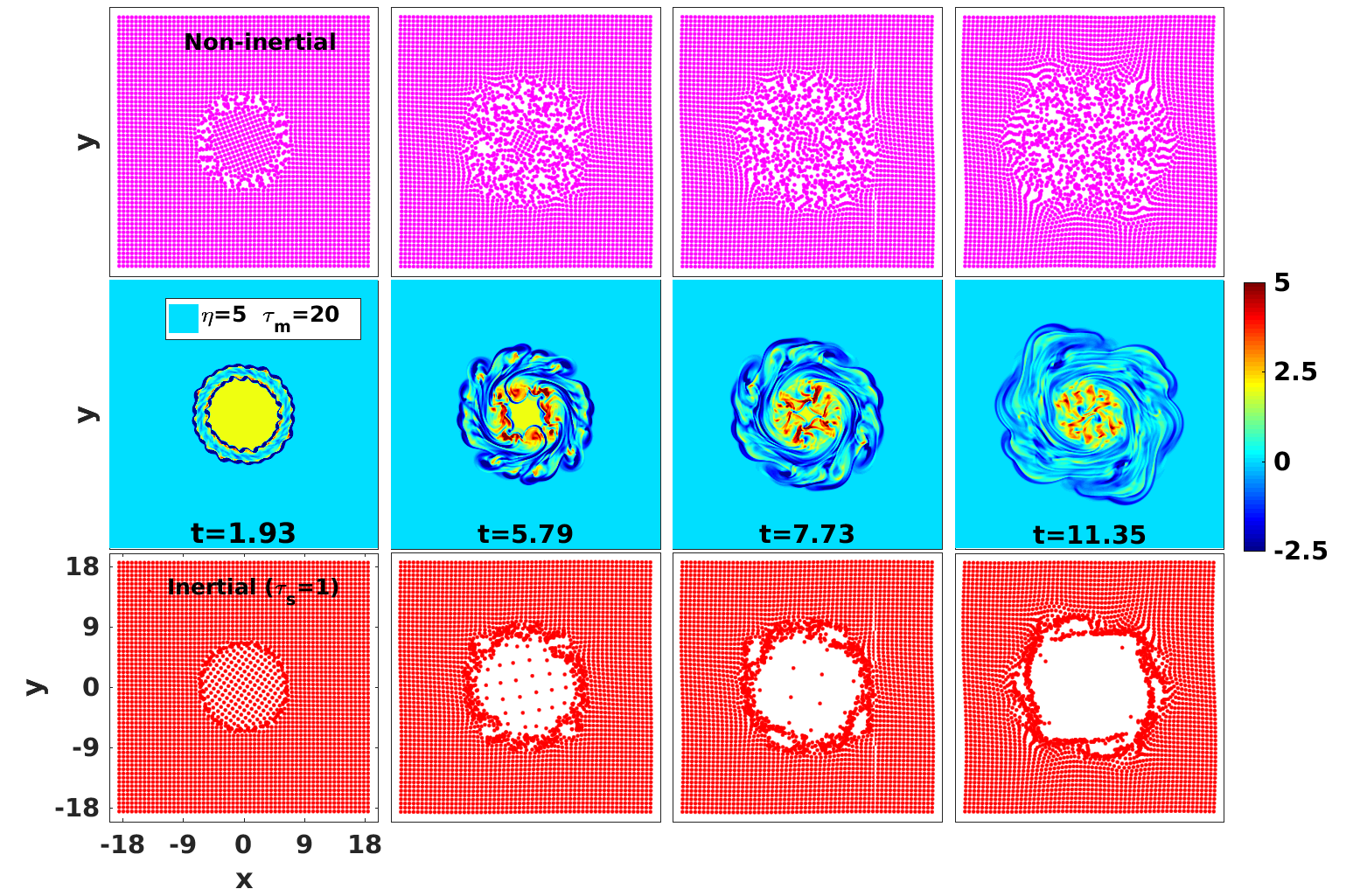}
	\caption {Viscoelastic fluid with coupling parameters ${\eta}=5$, and ${\tau_m}=20$ advect tracer particles. First and third row show the spatiotemporal distribution of non-inertial particles (shown by magenta dots) and inertial particles ($\tau_s$=1, shown by red dots), respectively, corresponding to the sharp vorticity profile evolution given in second row.} 
	\label{fig:st1_particle_sharp}
\end{figure} 
%\FloatBarrier
%~~~~~~~~~~~~~~~~~~~~~~~~~~~~~~~~~~~~~~~

 This accumulation process leads to the spatial inhomogeneous distribution of particles. This inhomogeneous distribution of the particles is known as clustering or preferential concentration.   Clustering is well-studied in the case of inertial particles \cite{maxey1987gravitational,squires1991preferential,fessler1994preferential,balkovsky2001intermittent,falkovich2002acceleration,goto2006self,petersen2019experimental,oka2021generalized}, and several studies are available for non-inertial particles \cite{drotos2019inhomogeneities}.
 
 In real flows the tracer particles always have some inertial value, so we compute the ensemble averaged MSD of intermediate inertial particles to analyze the diffusion of particles in the carrier VE fluid.  This diffusion of particles is associated with the mixing of the fluid. For this, we advect the inertial particles having $\tau_s$=1  using the sharp rotating flows discussed above,  inviscid fluid in Fig.~\ref{fig:sharpfluidphi1}/Fig.~\ref{fig:passive_particle_sharp_fluid}, VE fluid with $\eta=5$, $\tau_m=20$ in Fig.~\ref{fig:sharp520phi1}/Fig.~\ref{fig:st1_particle_sharp}, and another VE fluid with $\eta=2.5$, $\tau_m=20$ in Fig.~\ref{fig:sharp2p520phi1}. Initially t=0, we disperse 3600 tracer particles homogeneously over these fluids. 
 
%~~~~~~~~~~~~~~~~~~~~~~~~~~~~~~~~~~~~~~~
\begin{figure}
	\includegraphics[width=1.0\textwidth]{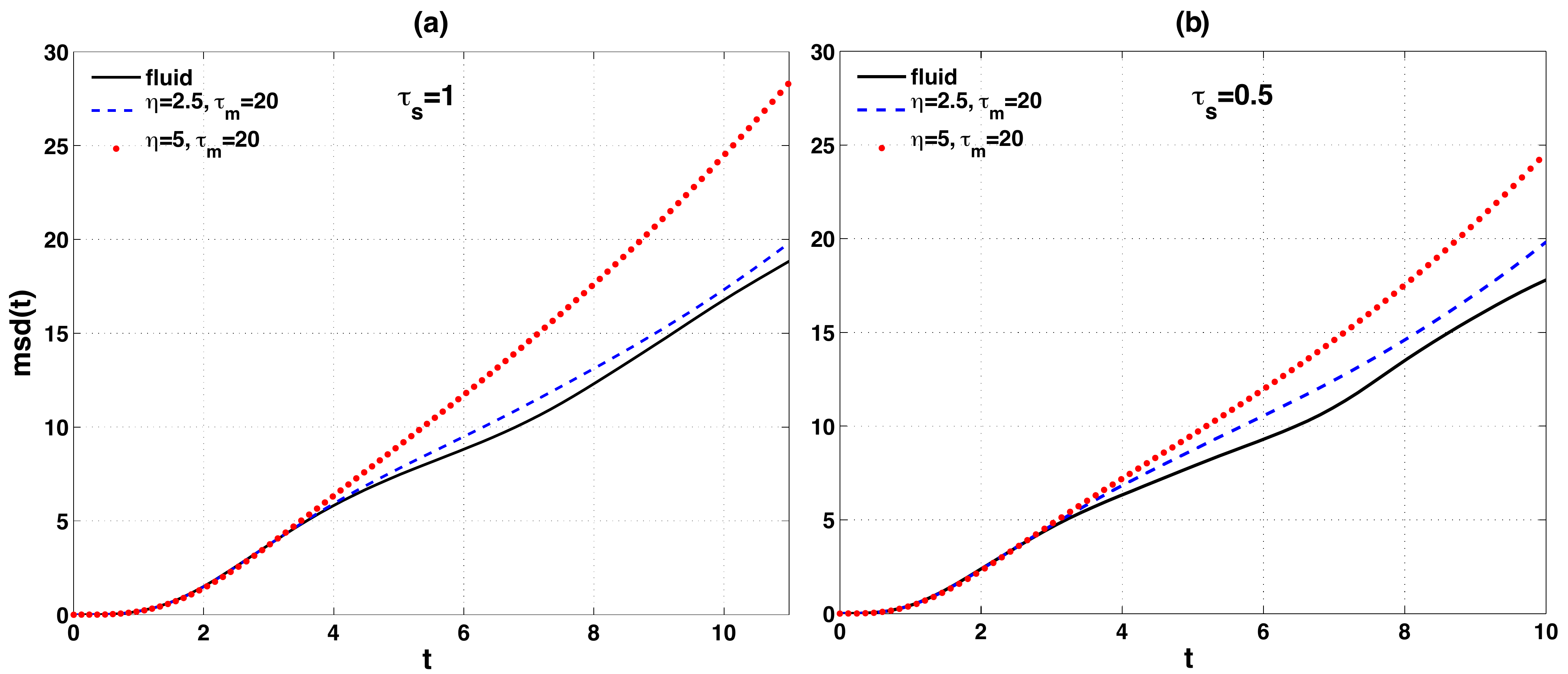}
	\caption{The MSD  as function of time for tracing particles. The sharp rotating vorticity advect the particles with (a) $\tau_s$=1, and (b) $\tau_s$=0.5. At later time, the MSD is proportional to the coupling strength.}
	\label{fig:sharp_msd_dcoeff_st1_fluid_2p520_520}
\end{figure}
%\FloatBarrier
%~~~~~~~~~~~~~~~~~~~~~~~~~~~~~~~~~~~~~~~

%The ratio of a particle time-scale to a fluid time-scale is known as Stokes number (St).
Figure~\ref{fig:sharp_msd_dcoeff_st1_fluid_2p520_520}(a) compares the ensemble averaged MSD values of inertial particles of $\tau_s$=1 for all three types of fluids. The evolution of MSD occurs in three stages. Initially (here, $0{\ge}t{\approx}1$), inertial  particles do not respond until their time-scale (here, $\tau_s=1$) is less than fluid time-scale or Stokes number (St) is less than unity. During the second stage the particles start to respond the flow  ($1{\ge}t{\le}3.8$), the slopes grow at the almost same rate for all fluids.  In the final stage ($t{\ge}3.8$), the slope shows the larger value for the higher coupling strength and it is minimal for inviscid fluid.  The slope of the MSD versus time is proportional to the diffusion coefficient of the tracer particles. Thus, the larger time-MSD slope means the carrier fluid shows a better mixing performance. Thus, this result is found consistent with our earlier observations made from the comparative analysis of pictorial evolution of vorticities in Figs. \ref{fig:sharpfluidphi1}, \ref{fig:sharp520phi1}, and \ref{fig:sharp2p520phi1}. In order to substantiate these observations we further compute the MSD of tracers with $\tau_s$=0.5 (Figure~\ref{fig:sharp_msd_dcoeff_st1_fluid_2p520_520}(b)) for the all three flows under the same flow conditions. From the plot, again the diffusion of particles increasing with coupling strength.  It should be noted that the MSD has been calculated up to time before hitting the flow to the boundaries. 

In the present work, although we have undertaken the simplest tracer model and preliminary investigation of tracers distribution (cluster formation) and their transport property (MSD), but it is worth attempting  as a foundation for the future study. 
 %~~~~~~~~~~~~~~~~~~~~~~~~~~~~~~~~~~~~~~~
 
%%%%%%%%%%%%%%%%%%%%%%%%%%%%%%%%%%%%%%%
%Observations and conclusions
%%%%%%%%%%%%%%%%%%%%%%%%%%%%%%%%%%%%%%%
\section{Summary and conclusions}
\label{discussion}
%~~~~~~~~~~~~~~~~~~~~~~~~
In this paper, we numerically explore the KH instability for a two-dimensional rotating SCDP under the formalism of GHD fluid model. This model treats the SCDP as a viscoelastic fluid. Here, we consider the specific cases of vorticity with abrupt radial changes, in particular: single-circulation, and multi-circulation vorticity shell profiles. We observe  the KH instability in form of small vortices at each interface along with a pair of ingoing and outgoing wavefronts of TS waves. The interactions between KH vortices, and with TS waves govern the mixing which has been quantified using the passive tracer particles simulation.  The advection of tracers with the flow has been noticed. Some main observations are as follows.

%~~~~~~~~~~~~~~~~~~~~~~~~
\begin{enumerate}
	\item 
	{
	The interplay between the TS waves and interacting KH vortices results in better mixing of VE fluids than standard hydrodynamic fluids where the only interaction between KH vortices happens. 
	}
\item 
{
 By tweaking the coupling strength parameter (ratio $\eta/\tau_m$) which usually represents the strength of viscoelasticity of the medium, one can control the evolution of KH instability that results in control over the transport properties like mixing and diffusion.
}
\item 
{
The GHD model especially developed for the study of SCDPs is extended to include the transport of passive Lagrangian  inertial and non-inertial tracer particles. This extended model is referred as Generalized Hydrodynamic Tracer Transport (GHTT) model. 
}
     \item 
     { 
     	We observe that the diffusion of intermediate inertial tracer particles in VE fluids is proportional to the coupling strength.  It is least for an inviscid fluid. 
      }
     \item 
     {The multi-circulation vortex profiles, we found that the relaxing rate of a turbulent medium increases with the increasing coupling strength.  
	}
\end{enumerate}
%~~~~~~~~~~~~~~~~~~~~~~~~
The particle tracking model appears a suitable diagnostic to understand the associated mixing in a fluid through the diffusion process, and nonlinear dust fluid dynamics through the clustering of a tracers. In present paper, we just discuss the preliminary simulations for tracers which are based on first order schemes like to advance the tracers we use first order RK scheme, and for the velocity the first order interpolation scheme.  Further improvisations of this model would be to include higher order schemes, finite size tracers, other forces e.g. gravity, inter-particle interactions, feedback from tracers to the  carrier fluid, etc.  These improvisations would be extremely fruitful which are left to future publications.

 Although, the rotational dust flows are extensively studied, but to our knowledge, no prior studies have explicitly examined the KH instability for the rotating flows, except ~\cite{dharodi2014visco, dharodi2020rotating}. An experimental research effort is required in order to valid this numerical work. 
 
%~~~~~~~~~~~~~~~~~~~~~~~~~~~~~~~~~~~~~~~~~
\newpage
\bibliographystyle{unsrt}
%\bibliography{vikram_mix}

\begin{thebibliography}{10}

\bibitem{drazin1970kelvin}
PG~Drazin.
\newblock Kelvin-helmholtz instability of finite amplitude.
\newblock {\em Journal of Fluid Mechanics}, 42(2):321--335, 1970.

\bibitem{chandrhd1981hydro}
S~Chandrasekhar.
\newblock Hydrodynamic and hydromagnetic stability, dover, new york, 1981.
\newblock 1981.

\bibitem{horton1987kelvin}
W~Horton, T~Tajima, and T~Kamimura.
\newblock Kelvin--helmholtz instability and vortices in magnetized plasma.
\newblock {\em The Physics of fluids}, 30(11):3485--3495, 1987.

\bibitem{van2010deep}
Hans van Haren and Louis Gostiaux.
\newblock A deep-ocean kelvin-helmholtz billow train.
\newblock {\em Geophysical Research Letters}, 37(3), 2010.

\bibitem{foullon2011magnetic}
Claire Foullon, Erwin Verwichte, Valery~M Nakariakov, Katariina Nykyri, and
  Charles~J Farrugia.
\newblock Magnetic kelvin-helmholtz instability at the sun.
\newblock {\em The Astrophysical Journal Letters}, 729(1):L8, 2011.

\bibitem{smyth2012ocean}
William~D Smyth and James~N Moum.
\newblock Ocean mixing by kelvin-helmholtz instability.
\newblock {\em Oceanography}, 25(2):140--149, 2012.

\bibitem{d1990kelvin}
N~d'Angelo and Bin Song.
\newblock The kelvin-helmholtz instability in dusty plasmas.
\newblock {\em Planetary and space science}, 38(12):1577--1579, 1990.

\bibitem{rawat1993kelvin}
SPS Rawat and NN~Rao.
\newblock Kelvin-helmholtz instability driven by sheared dust flow.
\newblock {\em Planetary and space science}, 41(2):137--140, 1993.

\bibitem{pandey2012shear}
BP~Pandey, SV~Vladimirov, and A~Samarian.
\newblock Shear driven instabilities in dusty plasmas.
\newblock In {\em Europhysics Conference on the Atomic and Molecular Physics of
  Ionized Gases (21st: 2012)}, pages 1--2. European Physical Society, 2012.

\bibitem{banerjee2012shear}
D~Banerjee, MS~Janaki, and N~Chakrabarti.
\newblock Shear flow instability in a strongly coupled dusty plasma.
\newblock {\em Physical Review E}, 85(6):066408, 2012.

\bibitem{dolai2016effect}
Bivash Dolai, RP~Prajapati, and RK~Chhajlani.
\newblock Effect of different dust flow velocities on combined kelvin-helmholtz
  and rayleigh-taylor instabilities in magnetized incompressible dusty fluids.
\newblock {\em Physics of Plasmas}, 23(11):113704, 2016.

\bibitem{ashwin2010kelvin}
J~Ashwin and R~Ganesh.
\newblock Kelvin helmholtz instability in strongly coupled yukawa liquids.
\newblock {\em Physical review letters}, 104(21):215003, 2010.

\bibitem{tiwari2012kelvina}
Sanat~Kumar Tiwari, Amita Das, Predhiman Kaw, and Abhijit Sen.
\newblock Kelvin--helmholtz instability in a weakly coupled dust fluid.
\newblock {\em Physics of Plasmas}, 19(2):023703, 2012.

\bibitem{tiwari2012kelvinb}
Sanat~Kumar Tiwari, Amita Das, Dilip Angom, Bhavesh~G Patel, and Predhiman Kaw.
\newblock Kelvin-helmholtz instability in a strongly coupled dusty plasma
  medium.
\newblock {\em Physics of Plasmas}, 19(7):073703, 2012.

\bibitem{tiwari2014kelvin}
Sanat Tiwari, Vikram Dharodi, Amita Das, Predhiman Kaw, Abhijit Sen, et~al.
\newblock Kelvin-helmholtz instability in dusty plasma medium: Fluid and
  particle approach.
\newblock {\em J. Plasma Phys}, 80(6):817--823, 2014.

\bibitem{luo2001kelvin}
QZ~Luo, N~D’Angelo, and RL~Merlino.
\newblock The kelvin--helmholtz instability in a plasma with negatively charged
  dust.
\newblock {\em Physics of Plasmas}, 8(1):31--35, 2001.

\bibitem{konopka2000rigid}
U~Konopka, D~Samsonov, AV~Ivlev, J~Goree, V~Steinberg, and GE~Morfill.
\newblock Rigid and differential plasma crystal rotation induced by magnetic
  fields.
\newblock {\em Physical Review E}, 61(2):1890, 2000.

\bibitem{klindworth2000laser}
M~Klindworth, A~Melzer, A~Piel, and VA~Schweigert.
\newblock Laser-excited intershell rotation of finite coulomb clusters in a
  dusty plasma.
\newblock {\em Physical Review B}, 61(12):8404, 2000.

\bibitem{schablinski2014sheared}
Jan Schablinski, Dietmar Block, Jan Carstensen, Franko Greiner, and Alexander
  Piel.
\newblock Sheared and unsheared rotation of driven dust clusters.
\newblock {\em Physics of Plasmas}, 21(7):073701, 2014.

\bibitem{choudhary2020three}
Mangilal Choudhary, Roman Bergert, Slobodan Mitic, and Markus~H Thoma.
\newblock Three-dimensional dusty plasma in a strong magnetic field:
  Observation of rotating dust tori.
\newblock {\em Physics of Plasmas}, 27(6):063701, 2020.

\bibitem{dharodi2014visco}
Vikram Singh~Dharodi, Sanat Kumar~Tiwari, and Amita Das.
\newblock Visco-elastic fluid simulations of coherent structures in strongly
  coupled dusty plasma medium.
\newblock {\em Physics of Plasmas}, 21(7):073705, 2014.

\bibitem{dharodi2020rotating}
Vikram~S Dharodi.
\newblock Rotating vortices in two-dimensional inhomogeneous strongly coupled
  dusty plasmas: Shear and spiral density waves.
\newblock {\em Physical Review E}, 102(4):043216, 2020.

\bibitem{kaw1998low}
PK~Kaw and A~Sen.
\newblock Low frequency modes in strongly coupled dusty plasmas.
\newblock {\em Physics of Plasmas}, 5(10):3552--3559, 1998.

\bibitem{dharodi2016sub}
Vikram Dharodi, Amita Das, Bhavesh Patel, and Predhiman Kaw.
\newblock Sub-and super-luminar propagation of structures satisfying
  poynting-like theorem for incompressible generalized hydrodynamic fluid model
  depicting strongly coupled dusty plasma medium.
\newblock {\em Physics of Plasmas}, 23(1):013707, 2016.

\bibitem{douady1991direct}
S~Douady, Y~Couder, and ME~Brachet.
\newblock Direct observation of the intermittency of intense vorticity
  filaments in turbulence.
\newblock {\em Physical review letters}, 67(8):983, 1991.

\bibitem{fessler1994preferential}
John~R Fessler, Jonathan~D Kulick, and John~K Eaton.
\newblock Preferential concentration of heavy particles in a turbulent channel
  flow.
\newblock {\em Physics of Fluids}, 6(11):3742--3749, 1994.

\bibitem{balkovsky2001intermittent}
E~Balkovsky, Gregory Falkovich, and A~Fouxon.
\newblock Intermittent distribution of inertial particles in turbulent flows.
\newblock {\em Physical Review Letters}, 86(13):2790, 2001.

\bibitem{falkovich2004intermittent}
Gregory Falkovich and Alain Pumir.
\newblock Intermittent distribution of heavy particles in a turbulent flow.
\newblock {\em Physics of Fluids}, 16(7):L47--L50, 2004.

\bibitem{zaichik2003two}
Leonid~I Zaichik, Olivier Simonin, and Vladimir~M Alipchenkov.
\newblock Two statistical models for predicting collision rates of inertial
  particles in homogeneous isotropic turbulence.
\newblock {\em Physics of Fluids}, 15(10):2995--3005, 2003.

\bibitem{bec2006acceleration}
Jeremy Bec, Luca Biferale, Guido Boffetta, Antonio Celani, Massimo Cencini,
  Alessandra Lanotte, S~Musacchio, and Federico Toschi.
\newblock Acceleration statistics of heavy particles in turbulence.
\newblock {\em Journal of Fluid Mechanics}, 550:349--358, 2006.

\bibitem{squires1991preferential}
Kyle~D Squires and John~K Eaton.
\newblock Preferential concentration of particles by turbulence.
\newblock {\em Physics of Fluids A: Fluid Dynamics}, 3(5):1169--1178, 1991.

\bibitem{boivin1998direct}
Marc Boivin, Olivier Simonin, and Kyle~D Squires.
\newblock Direct numerical simulation of turbulence modulation by particles in
  isotropic turbulence.
\newblock {\em Journal of Fluid Mechanics}, 375:235--263, 1998.

\bibitem{reade2000numerical}
Walter~C Reade and Lance~R Collins.
\newblock A numerical study of the particle size distribution of an aerosol
  undergoing turbulent coagulation.
\newblock {\em Journal of Fluid Mechanics}, 415:45--64, 2000.

\bibitem{zhou2001modelling}
Yong Zhou, Anthony~S Wexler, and Lian-Ping Wang.
\newblock Modelling turbulent collision of bidisperse inertial particles.
\newblock {\em Journal of Fluid Mechanics}, 433:77, 2001.

\bibitem{yeung2001lagrangian}
PK~Yeung.
\newblock Lagrangian characteristics of turbulence and scalar transport in
  direct numerical simulations.
\newblock {\em Journal of fluid mechanics}, 427:241, 2001.

\bibitem{yeung2002lagrangian}
PK~Yeung.
\newblock Lagrangian investigations of turbulence.
\newblock {\em Annual review of fluid mechanics}, 34(1):115--142, 2002.

\bibitem{ishihara2002relative}
Takashi Ishihara and Yukio Kaneda.
\newblock Relative diffusion of a pair of fluid particles in the inertial
  subrange of turbulence.
\newblock {\em Physics of Fluids}, 14(11):L69--L72, 2002.

\bibitem{collins2004reynolds}
Lance~R Collins and Arun Keswani.
\newblock Reynolds number scaling of particle clustering in turbulent aerosols.
\newblock {\em New Journal of Physics}, 6(1):119, 2004.

\bibitem{biferale2004multifractal}
L~Biferale, Guido Boffetta, Antonio Celani, BJ~Devenish, Alessandra Lanotte,
  and Federico Toschi.
\newblock Multifractal statistics of lagrangian velocity and acceleration in
  turbulence.
\newblock {\em Physical review letters}, 93(6):064502, 2004.

\bibitem{chun2005clustering}
Jaehun Chun, Donald~L Koch, Sarma~L Rani, Aruj Ahluwalia, and Lance~R Collins.
\newblock Clustering of aerosol particles in isotropic turbulence.
\newblock {\em Journal of Fluid Mechanics}, 536:219, 2005.

\bibitem{biferale2005particle}
Luca Biferale, Guido Boffetta, Antonio Celani, Alessandra Lanotte, and Federico
  Toschi.
\newblock Particle trapping in three-dimensional fully developed turbulence.
\newblock {\em Physics of Fluids}, 17(2):021701, 2005.

\bibitem{cencini2006dynamics}
M~Cencini, J~Bec, L~Biferale, Guido Boffetta, A~Celani, AS~Lanotte,
  S~Musacchio, and F~Toschi.
\newblock Dynamics and statistics of heavy particles in turbulent flows.
\newblock {\em Journal of Turbulence}, (7):N36, 2006.

\bibitem{ott2000experimental}
S{\o}ren Ott and Jakob Mann.
\newblock An experimental investigation of the relative diffusion of particle
  pairs in three-dimensional turbulent flow.
\newblock {\em Journal of Fluid Mechanics}, 422:207--223, 2000.

\bibitem{la2001fluid}
Arthur La~Porta, Greg~A Voth, Alice~M Crawford, Jim Alexander, and Eberhard
  Bodenschatz.
\newblock Fluid particle accelerations in fully developed turbulence.
\newblock {\em Nature}, 409(6823):1017--1019, 2001.

\bibitem{mordant2001measurement}
Nicolas Mordant, Pascal Metz, Olivier Michel, and J-F Pinton.
\newblock Measurement of lagrangian velocity in fully developed turbulence.
\newblock {\em Physical Review Letters}, 87(21):214501, 2001.

\bibitem{voth2002measurement}
Greg~A Voth, Arthur La~Porta, Alice~M Crawford, Jim Alexander, and Eberhard
  Bodenschatz.
\newblock Measurement of particle accelerations in fully developed turbulence.
\newblock {\em Journal of Fluid Mechanics}, 469:121--160, 2002.

\bibitem{sawford2003conditional}
Brian~Lewis Sawford, PK~Yeung, Michael~S Borgas, Prakash Vedula, Arthur
  La~Porta, Alice~M Crawford, and Eberhard Bodenschatz.
\newblock Conditional and unconditional acceleration statistics in turbulence.
\newblock {\em Physics of Fluids}, 15(11):3478--3489, 2003.

\bibitem{mason1997particle}
Thomas~G Mason, Kavita Ganesan, J~Harry van Zanten, Denis Wirtz, and Scot~C
  Kuo.
\newblock Particle tracking microrheology of complex fluids.
\newblock {\em Physical review letters}, 79(17):3282, 1997.

\bibitem{waigh2005microrheology}
Thomas~A Waigh.
\newblock Microrheology of complex fluids.
\newblock {\em Reports on progress in physics}, 68(3):685, 2005.

\bibitem{falkovich2001particles}
Gregory Falkovich, K~Gawedzki, and Massimo Vergassola.
\newblock Particles and fields in fluid turbulence.
\newblock {\em Reviews of modern Physics}, 73(4):913, 2001.

\bibitem{schwabe2014collective}
Mierk Schwabe, Sergey Zhdanov, Christoph R{\"a}th, David~B Graves, Hubertus~M
  Thomas, and Gregor~E Morfill.
\newblock Collective effects in vortex movements in complex plasmas.
\newblock {\em Physical review letters}, 112(11):115002, 2014.

\bibitem{riley1974diffusion}
James~J Riley and GS~Patterson~Jr.
\newblock Diffusion experiments with numerically integrated isotropic
  turbulence.
\newblock {\em The Physics of Fluids}, 17(2):292--297, 1974.

\bibitem{mclaughlin1989aerosol}
John~B McLaughlin.
\newblock Aerosol particle deposition in numerically simulated channel flow.
\newblock {\em Physics of Fluids A: Fluid Dynamics}, 1(7):1211--1224, 1989.

\bibitem{tiwari2012longitudinal}
Sanat~Kumar Tiwari, Amita Das, Predhiman Kaw, and Abhijit Sen.
\newblock Longitudinal singular response of dusty plasma medium in weak and
  strong coupling limits.
\newblock {\em Physics of Plasmas}, 19(1):013706, 2012.

\bibitem{sapsis2010clustering}
Themistoklis Sapsis and George Haller.
\newblock Clustering criterion for inertial particles in two-dimensional
  time-periodic and three-dimensional steady flows.
\newblock {\em Chaos: An Interdisciplinary Journal of Nonlinear Science},
  20(1):017515, 2010.

\bibitem{maxey1983equation}
Martin~R Maxey and James~J Riley.
\newblock Equation of motion for a small rigid sphere in a nonuniform flow.
\newblock {\em The Physics of Fluids}, 26(4):883--889, 1983.

\bibitem{guha2008transport}
Abhijit Guha.
\newblock Transport and deposition of particles in turbulent and laminar flow.
\newblock {\em Annu. Rev. Fluid Mech.}, 40:311--341, 2008.

\bibitem{jeon2013anomalous}
Jae-Hyung Jeon, Natascha Leijnse, Lene~B Oddershede, and Ralf Metzler.
\newblock Anomalous diffusion and power-law relaxation of the time averaged
  mean squared displacement in worm-like micellar solutions.
\newblock {\em New Journal of Physics}, 15(4):045011, 2013.

\bibitem{boris_book}
J.~P. Boris, A.~M. Landsberg, E.~S. Oran, and J.~H. Gardner.
\newblock {\em LCPFCT A flux-corrected transport algorithm for solving
  generalized continuity equations}.
\newblock Technical Report NRL Memorandum Report 93-7192, Naval Research
  Laboratory, 1993.

\bibitem{swarztrauber1999fishpack}
P~Swarztrauber, R~Sweet, and John~C Adams.
\newblock Fishpack: Efficient fortran subprograms for the solution of elliptic
  partial differential equations.
\newblock {\em UCAR Publication, July}, 1999.

\bibitem{ravichandran2017clustering}
S~Ravichandran, P~Deepu, and Rama Govindarajan.
\newblock Clustering of heavy particles in vortical flows: a selective review.
\newblock {\em S{\=a}dhan{\=a}}, 42(4):597--605, 2017.

\bibitem{maxey1987gravitational}
Martin~R Maxey.
\newblock The gravitational settling of aerosol particles in homogeneous
  turbulence and random flow fields.
\newblock {\em Journal of fluid mechanics}, 174:441--465, 1987.

\bibitem{falkovich2002acceleration}
G~Falkovich, A~Fouxon, and MG~Stepanov.
\newblock Acceleration of rain initiation by cloud turbulence.
\newblock {\em Nature}, 419(6903):151--154, 2002.

\bibitem{goto2006self}
Susumu Goto and JC~Vassilicos.
\newblock Self-similar clustering of inertial particles and zero-acceleration
  points in fully developed two-dimensional turbulence.
\newblock {\em Physics of Fluids}, 18(11):115103, 2006.

\bibitem{petersen2019experimental}
Alec~J Petersen, Lucia Baker, and Filippo Coletti.
\newblock Experimental study of inertial particles clustering and settling in
  homogeneous turbulence.
\newblock {\em Journal of Fluid Mechanics}, 864:925--970, 2019.

\bibitem{oka2021generalized}
Sunao Oka and Susumu Goto.
\newblock Generalized sweep-stick mechanism of inertial-particle clustering in
  turbulence.
\newblock {\em Physical Review Fluids}, 6(4):044605, 2021.

\bibitem{drotos2019inhomogeneities}
G{\'a}bor Dr{\'o}tos, Pedro Monroy, Emilio Hern{\'a}ndez-Garc{\'\i}a, and
  Crist{\'o}bal L{\'o}pez.
\newblock Inhomogeneities and caustics in the sedimentation of noninertial
  particles in incompressible flows.
\newblock {\em Chaos: An Interdisciplinary Journal of Nonlinear Science},
  29(1):013115, 2019.

\end{thebibliography}

%~~~~~~~~~~~~~~~~~~~~~~~~~
\end{document}